\documentclass{aastex62}
\pdfoutput=1
\usepackage{amsmath,amssymb,CJK,soul}
\usepackage{longtable}
\usepackage{graphicx}
\usepackage{lineno}
\received{February 22, 2021}
\revised{March 17, 2021}
\accepted{March 26, 2021}
\submitjournal{Astrophysical Journal Letters}

\shorttitle{Photometric Lightcurves of inactive (6478) Gault}
\shortauthors{Purdum et al.}

\begin{document}
	
\title{Time-series and Phasecurve Photometry of Episodically-Active Asteroid (6478) Gault in a Quiescent State Using APO, GROWTH, P200 and ZTF}

\correspondingauthor{Josiah Purdum}
\email{jnpurdum@gmail.com}

\author{Josiah N. Purdum}
\affiliation{Department of Astronomy, San Diego State University, 5500 Campanile Dr, San Diego, CA 92182, U.S.A.}

\author{Zhong-Yi Lin$^*$}
\affiliation{Institute of Astronomy, National Central University, Taoyuan 32001, Taiwan}
\altaffiliation{These authors contributed equally to this work.}

\author[0000-0002-4950-6323]{Bryce T. Bolin$^*$}
\affiliation{IPAC, Mail Code 100-22, Caltech, 1200 E. California Blvd., Pasadena, CA 91125, U.S.A.}
\altaffiliation{These authors contributed equally to this work.}

\author[0000-0002-4477-3625]{Kritti Sharma}
\affiliation{Department of Mechanical Engineering, Indian Institute of Technology Bombay, Powai, Mumbai-400076, India}

\author{Philip I. Choi}
\affiliation{Physics and Astronomy Department, Pomona College, 333 N. College Way, Claremont, CA 91711, U.S.A.}

\author[0000-0002-6112-7609]{Varun Bhalerao}
\affiliation{Department of Physics, Indian Institute of Technology Bombay, Powai, Mumbai-400076, India}

\author[0000-0002-2934-3723]{Josef Hanu\v{s}}
\affiliation{Institute of Astronomy, Faculty of Mathematics and Physics, Charles University, V~Hole{\v s}ovi{\v c}k{\'a}ch 2, 18000 Prague, Czech Republic}

\author[0000-0003-0871-4641]{Harsh Kumar}
\affiliation{Department of Physics, Indian Institute of Technology Bombay, Powai, Mumbai-400076, India}
\affiliation{LSSTC Data Science Fellow}

\author[0000-0001-9171-5236]{Robert Quimby}
\affiliation{Department of Astronomy, San Diego State University, 5500 Campanile Dr, San Diego, CA 92182, U.S.A.}
\affiliation{Kavli Institute for the Physics and Mathematics of the Universe (WPI), The University of Tokyo Institutes for Advanced Study, The University of Tokyo, Kashiwa, Chiba 277-8583, Japan}

\author{Joannes C. Van Roestel}
\affiliation{Division of Physics, Mathematics and Astronomy, California Institute of Technology, Pasadena, CA 91125, U.S.A.}

\author{Chengxing Zhai}
\affiliation{Jet Propulsion Laboratory, California Institute of Technology, 4800 Oak Grove Drive, Pasadena, CA 91109, U.S.A.}

\author[0000-0003-1156-9721]{Yanga R. Fernandez}
\affiliation{Dept. of Physics and Florida Space Inst., Univ. of Central Florida, 4000 Central Florida Boulevard, Orlando FL 32816-2385, USA}

\author[0000-0002-9548-1526]{Carey M. Lisse}
\affiliation{Johns Hopkins University Applied Physics Laboratory, Laurel, MD 20723}

\author[0000-0002-2668-7248]{Dennis Bodewits}
\affiliation{Physics Department, Leach Science Center, Auburn University, Auburn, AL 36832, U.S.A.}

\author{Christoffer Fremling}
\affiliation{Division of Physics, Mathematics and Astronomy, California Institute of Technology, Pasadena, CA 91125, U.S.A.}

\author[0000-0003-2632-572X]{Nathan Ryan Golovich}
\affiliation{Lawrence Livermore National Laboratory, 7000 East Avenue, Livermore, CA 94550, U.S.A.}

\author{Chen-Yen Hsu}
\affiliation{Graduate Institute of Astronomy, National Central University, 32001, Taiwan}

\author{Wing-Huen Ip}
\affiliation{Institute of Astronomy, National Central University, 32001, Taiwan}

\author[0000-0001-8771-7554]{Chow-Choong Ngeow}
\affiliation{Graduate Institute of Astronomy, National Central University, 32001, Taiwan}

\author{Navtej S. Saini}
\affiliation{Jet Propulsion Laboratory, California Institute of Technology, 4800 Oak Grove Drive, Pasadena, CA 91109, U.S.A.}

\author{Michael Shao}
\affiliation{Jet Propulsion Laboratory, California Institute of Technology, 4800 Oak Grove Drive, Pasadena, CA 91109, U.S.A.}

\author{Yuhan Yao}
\affiliation{Division of Physics, Mathematics and Astronomy, California Institute of Technology, Pasadena, CA 91125, U.S.A.}

\author{Tom\'{a}s Ahumada}
\affiliation{Department of Astronomy, University of Maryland, College Park, MD 20742, U.S.A.}

\author{Shreya Anand}
\affiliation{Division of Physics, Mathematics and Astronomy, California Institute of Technology, Pasadena, CA 91125, U.S.A.}

\author{Igor Andreoni}
\affiliation{Division of Physics, Mathematics and Astronomy, California Institute of Technology, Pasadena, CA 91125, U.S.A.}

\author{Kevin B. Burdge}
\affiliation{Division of Physics, Mathematics and Astronomy, California Institute of Technology, Pasadena, CA 91125, U.S.A.}

\author{Rick Burruss}
\affiliation{Caltech Optical Observatories, California Institute of Technology, Pasadena, CA 91125, U.S.A.}

\author[0000-0003-1656-4540]{Chan-Kao Chang}
\affiliation{Graduate Institute of Astronomy, National Central University, 32001, Taiwan}

\author{Chris M. Copperwheat}
\affiliation{Astrophysics Research Institute Liverpool John Moores University, 146 Brownlow Hill, Liverpool L3 5RF, United Kingdom}

\author{Michael Coughlin}
\affil{School of Physics and Astronomy, University of Minnesota, Minneapolis, Minnesota 55455, USA}

\author{Kishalay De}
\affiliation{Division of Physics, Mathematics and Astronomy, California Institute of Technology, Pasadena, CA 91125, U.S.A.}

\author[0000-0002-5884-7867]{Richard Dekany}
\affiliation{Caltech Optical Observatories, California Institute of Technology, Pasadena, CA 91125, U.S.A.}

\author{Alexandre Delacroix}
\affiliation{Caltech Optical Observatories, California Institute of Technology, Pasadena, CA 91125, U.S.A.}

\author{Andrew Drake}
\affiliation{Division of Physics, Mathematics and Astronomy, California Institute of Technology, Pasadena, CA 91125, USA}

\author[0000-0001-5060-8733]{Dmitry Duev}
\affiliation{Division of Physics, Mathematics and Astronomy, California Institute of Technology, Pasadena, CA 91125, U.S.A.}

\author{Matthew Graham}
\affiliation{Division of Physics, Mathematics and Astronomy, California Institute of Technology, Pasadena, CA 91125, U.S.A.}

\author{David Hale}
\affiliation{Caltech Optical Observatories, California Institute of Technology, Pasadena, CA 91125, U.S.A.}

\author{Erik C. Kool}
\affiliation{DivisionThe Oskar Klein Centre, Department of Astronomy, Stockholm University, AlbaNova, SE-10691, Stockholm, Sweden}
\affiliation{Department of Physics and Astronomy, Macquarie University, NSW 2109, Sydney, Australia}

\author{Mansi M. Kasliwal}
\affiliation{Division of Physics, Mathematics and Astronomy, California Institute of Technology, Pasadena, CA 91125, U.S.A.}

\author{Iva S. Kostadinova}
\affiliation{Division of Physics, Mathematics and Astronomy, California Institute of Technology, Pasadena, CA 91125, U.S.A.}

\author{Shrinivas R. Kulkarni}
\affiliation{Division of Physics, Mathematics and Astronomy, California Institute of Technology, Pasadena, CA 91125, U.S.A.}

\author[0000-0003-2451-5482]{Russ R. Laher}
\affiliation{IPAC, Mail Code 100-22, Caltech, 1200 E. California Blvd., Pasadena, CA 91125, USA}

\author[0000-0003-2242-0244]{Ashish Mahabal}
\affiliation{Division of Physics, Mathematics and Astronomy, California Institute of Technology, Pasadena, CA 91125, U.S.A.}
\affiliation{Center for Data Driven Discovery, California Institute of Technology, Pasadena, CA 91125, U.S.A.}

\author{Frank J. Masci}
\affiliation{IPAC, Mail Code 100-22, Caltech, 1200 E. California Blvd., Pasadena, CA 91125, USA}

\author{Przemyslaw J. Mr\'{o}z}
\affiliation{Division of Physics, Mathematics and Astronomy, California Institute of Technology, Pasadena, CA 91125, U.S.A.}

\author{James D. Neill}
\affiliation{Division of Physics, Mathematics and Astronomy, California Institute of Technology, Pasadena, CA 91125, U.S.A.}

%\author{Thomas A. Prince}
%\affiliation{Division of Physics, Mathematics and Astronomy, California Institute of Technology, Pasadena, CA 91125, U.S.A.}

\author[0000-0002-0387-370X]{Reed Riddle}
\affiliation{Caltech Optical Observatories, California Institute of Technology, Pasadena, CA 91125, U.S.A.}

\author{Hector Rodriguez}
\affiliation{Caltech Optical Observatories, California Institute of Technology, Pasadena, CA 91125, U.S.A.}

\author[0000-0001-7062-9726]{Roger M. Smith}
\affiliation{Caltech Optical Observatories, California Institute of Technology, Pasadena, CA 91125, U.S.A.}

\author{Richard Walters}
\affiliation{Division of Physics, Mathematics and Astronomy, California Institute of Technology, Pasadena, CA 91125, U.S.A.}

\author[0000-0003-1710-9339]{Lin Yan}
\affil{Caltech Optical Observatories, California Institute of Technology, Pasadena, CA 91125, U.S.A.}

\author{Jeffry Zolkower}
\affiliation{Caltech Optical Observatories, California Institute of Technology, Pasadena, CA 91125, U.S.A.}

\begin{abstract}
	We observed Episodically Active Asteroid (6478) Gault in 2020 with multiple telescopes in Asia and North America and have found that it is no longer active after its recent outbursts at the end of 2018 and start of 2019. The inactivity during this apparation allowed us to measure the absolute magnitude of Gault of $H_r$ = 14.63$\pm$0.02, $G_r$ = 0.21$\pm$0.02 from our secular phasecurve observations. In addition, we were able to constrain Gault's rotation period using time-series photometric lightcurves taken over 17 h on multiple days in 2020 August, September and October. The photometric lightcurves have a repeating $\lesssim$0.05 magnitude feature suggesting that (6478) Gault has a rotation period of $\sim$2.5 hours and may have a semi-spherical or top-like shape, much like Near-Earth Asteroids Ryugu and Bennu. The rotation period of $\sim$2.5 hours is near to the expected critical rotation period for an asteroid with the physical properties of (6478) Gault suggesting that its activity observed over multiple epochs is due to surface mass shedding from its fast rotation spun up by the Yarkovsky–O'Keefe–Radzievskii–Paddack effect.
\end{abstract}

\section{Introduction}
Active asteroids produce comet-like tails and comae that can be driven by many different types of forces different from comets themselves \citep{Jewitt_2015}. While sublimation of water ice is a primary driver for activity in `typical' comets, the $\sim$20 known (so far) active asteroids in the Main Belt seem to lose mass via a wider array of physical effects such as collisions \citep[e.g.,][]{Snodgrass_2010}, rotational instabilities \citep[e.g.,][]{Jewitt_2013}, and thermal fracture \citep[e.g.,][]{Jewitt_2019}. We can assess the physics of a particular active asteroid's activity via observations over long time baselines that assess the object's photometric and morphological development. As more and more active asteroids are discovered, it is vital to continuously monitor these objects and determine the frequency of the various phenomena in the Main Belt.

Main Belt asteroid (6478) Gault (1998 JC$_1$; ``Gault" hereafter) has been the subject of wide interest since the discovery in early 2019 of comet tail-like extended emission \citep{Smith_2019}. Eventually three tails were noted in January 2019 \citep{Ye_2019, Jewitt_2019b} on the S-type Phocaea family member \citep{Sanchez_2019}, suggesting multiple sporadic outbursts of activity. \cite{Ye_2019} assessed the dynamics of the dust seen near Gault and estimated that two outbursts had actually occurred in late 2018. Searches through archival data that serendipitously caught Gault  revealed that there had been active episodes in 2013, 2016, and 2017 as well \citep{Chandler_2019}.

Many authors have proposed that the cause of Gault's activity is the instability of material on its surface \citep{,Hui_2019, Jewitt_2019b, Kleyna_2019, Moreno_2019,  Ye_2019} due to its rotation being spun up by the the Yarkovsky-O’Keefe-Radzievskii-Paddack (YORP) effect  \citep{Bottke_2006, Kleyna_2019}. A critical observational test of this hypothesis would be to measure the asteroid's rotational period. Unfortunately, due to the dust around the asteroid, the rotational period had not been well constrained as reflected sunlight from dust grains would swamp the signal from the asteroid itself and thus suppress short-term lightcurve variations due to Gault's shape, as indeed \cite{Jewitt_2019b} concluded. However there have been published reports of some hints of rotational signatures in lightcurve data. For example, \cite{Ivanova_2020} suggest a rotation period of 1.79 hours, \cite{Carbognani_Buzzoni_2020} suggest 3.34 hours, \cite{Ferrin_2019} have 3.36 hours, and \cite{Kleyna_2019} suggest $\sim$2 hours. In all cases, the lightcurve amplitude was quite small, on the order of just a few hundredths of a magnitude, which would be on the same order of the signal noise. It should also be noted that others \citep[e.g.,][]{Moreno_2019} report no variation in photometry over a time span longer than these periods. This very small amplitude demonstrates the challenge of photometrically extracting a rotation period from an active body \citep[see also][ for an example]{Bolin2020HST}.

In this paper we report on several sets of imaging and photometry of Gault obtained in 2019 while it was still active as well as in 2020 when the asteroid appeared to be quiescent \citep{Purdum_2020}. These datasets have allowed us to constrain the rotation period of Gault. We also use all of the data to understand the longer-term, secular variations in Gault's activity. In \S 2 we describe the observations from 2019 and 2020. In \S 3 we present the photometry, and in \S 4 we discuss Gault's behavior, spin state, and shape, and summarize the article.

\section{Observations}
For our analysis we have made use of both our own, PI-led, pointed observations using several telescope facilities in the GROWTH (Global Relay of Observatories Watching Transients Happen) network \citep[][]{Kasliwal2019} and other faculties, as well as archival data from the Zwicky Transient Survey \citep[ZTF,][]{Graham_2019, Bellm_2019}. Our pointed observations occurred on 17 nights between 2019 January 8 and 2020 October 20, and made use of six telescopes. On 7 of those 17 nights, we were able to have multiple telescopes follow Gault in a coordinated effort. The observations in 2019 showed Gault to still be active, but all such observations in 2020 showed no activity, only a point source \citep{Purdum_2020}. The ZTF data are from 89 nights from 2018 November 01 to 2020 October 14. All data in this work have been flat- and bias-corrected since only CCDs were used. Tables \ref{table1} and \ref{table2} list the technical details of each telescope and the particulars about each observing run, respectively, while Tables \ref{table3} and \ref{table4} are the photometric data plotted for this work and the archival ZTF survey, respectively. We describe below each telescope facility used in our work.

\subsection{Mount Laguna Observatory 1.0-Meter Telescope}
Images of Gault were taken with the Mount Laguna Observatory (MLO) 1.0-meter telescope on 2020 June 24 UT, several months after the asteroid was leaving Solar conjunction. 120 images were taken with 30-second exposure times each, culminating 60 minutes of total exposure to measure the morphology of the asteroid.

Later observations were taken of Gault with MLO with an aim to constrain a rotation period. These data were taken between 2020 August 23 UT and 2020 October 20 UT over 6 separate campaigns listed in Table \ref{table2}. MLO lightcurve images were taken in the Johnson-Cousins R filter with between 165 and 190 separate 120-second exposures.  

\subsection{Lulin One-meter Telescope}
The time-series observations of Gault using the Lulin one-meter Telescope (LOT) at the Lulin Observatory, Taiwan, for 2020 August 23 and 24 UT lasted 6.4 hours and 5.2 hours, respectively. The other time span on 2020 September 21 UT and 2020 October 11 UT lasted 6.4 hours and 6.8 hours, respectively. Except for the use of unfiltered CCD observations early in the campaign, all observations are acquired with R-filter and obtained through non-sidereal tracking. 

\subsection{Astrophysics Research Consortium 3.5-meter Telescope}
Between 2019 January 8 UT and 2019 June 6 UT we observed Gault over 6 campaigns with the Astrophysics Research Consortium 3.5-meter Telescope (ARC) at Apache Point Observatory before the asteroid headed into Solar conjunction (Table \ref{table2}). Individual observations created lightcurves spanning between 1 hour (2019 March 24 UT) and 4 hours (2019 April 26 UT). Observations were taken with the ARCTIC optical CCD \citep{Huehnerhoff_2016} in the $r'$ filter with an average seeing of 1.8\arcsec. Throughout January and February of 2019, Gault remained visibly active, while observations between March and June of 2019 displayed just a remnant tail (see Figures \ref{Active} and \ref{ARC_lightcurve}).

\subsection{Palomar Observatory 200-Inch Telescope} 
On 2020 Aug 27 UT, the Palomar Observatory 200-inch telescope (P200) observed Gault in the \textit{r} band. The 103 exposures each had an equivalent exposure time of 90 seconds, accumulating in a total of 9,270 seconds with an average seeing of $\sim$1.5\arcsec. Observations were made with the Caltech HIgh-speed Multi-color camERA \citep[CHIMERA,][]{Harding_2016}. Although CHIMERA observes in two simultaneous optical bands, we only use the \textit{r} band in this work.

\subsection{Zwicky Transient Facility using the Palomar 48-Inch Telescope}
Images of Gault were taken with the Zwicky Transient Facility (ZTF) \citep{Graham_2019} which is mounted on the Palomar Observatory's 48-inch telescope (P48) \citep{Bellm_2019,Dekany2020}. ZTF images are located in the ZTF archive \citep{Masci2019} and Gault's photometry was measured with a 5\arcsec~radius aperture and processed using the ZChecker software \citep{Kelley_2019}. The 30-second exposure time observations were made in the \textit{r} band and color-corrected using the $g-r$ value of 0.50 $\pm$ 0.04 from \cite{Ye_2019}. Data in Figure \ref{ZTF_Phase} between 2018 November 1 UT and 2019 February 10 UT are adapted from \cite{Ye_2019}. A full list of the ZTF observations starting 2020 April 2 UT and ending 2020 October 14 UT is located in Table \ref{table4}. The seeing varied between 1.5\arcsec--2.5\arcsec~and the airmasses varied from 1.4 to 2.6 during the span of our observations.

\subsection{GROWTH-India Telescope}
We observed Gault on multiple nights with the 0.7m GROWTH-India telescope (GIT), using the SDSS $r^\prime$ filter and an Apogee KAF3200EB camera giving a $\sim 11^\prime \times 7.5^\prime $ field of view. Due to the slow motion of Gault, we used sidereal tracking and took multiple exposures. Data were acquired in 120~sec exposures on 2020 September 21, followed by 180~sec exposures on 2020 October 16 and October 20. 
% GIT data processing start
Data were downloaded and processed in real time at our data processing machine at IIT Bombay. We calibrated images for processing by applying bias correction and flat-fielding, obtained an astrometry solution using the offline engine of astrometry.net \citep{Lang_2010} and finally removed cosmic-rays via Astro-SCRAPPY~\citep{2019ascl.soft07032M} package. Photometry was performed using PyRAF based processing pipeline. We cross-matched the the sextractor-identified~\citep{bertin11} sources in the GIT image with \#~II/349/ps1 catalogue~\citep{2018AAS...23143601F} using vizier. Magnitudes were calibrated by correcting for zero points.
% GIT data processing end

\subsection{GROWTH Coordinated Observations}
On 2020 September 21 UT, LOT, GIT, MLO, and the Table Mountain Observatory (TMO) 1.0-meter telescope participated in a 20-hour relay of observing Gault for photometric lightcurve variation. MLO started the relay at 2020-09-21 04:24:47 UT (airmass 2.0) and ran continuous observations of 120-second exposure times until 11:35:27 UT (airmass 2.0) the same day, totaling 22800 seconds of exposure time. LOT took over shortly after at 14:26:07 (airmass 1.2) amassing 14130 seconds of 90-second exposures before finishing at 20:29:09 UT (airmass 2.5). GIT observed Gault for roughly 6 hours in 135 images amassing 16,200 seconds of exposure starting from 17:21:31 UT (airmass 1.2), in the middle of the LOT observations, and ending at 23:18:07 UT (airmass 2.5). TMO observed Gault for roughly 4 hours in 225 images amassing 13,500 seconds of exposure starting at 06:36:37 UT (airmass 1.3), in the middle of MLO's observations. Typical seeing for GIT is 2.5\arcsec~and for TMO is 1.5\arcsec. We were able to take data with Table Mountain Observatory despite the degraded conditions caused by a nearby wildfire. Additional coordination among the GROWTH network include MLO-LOT observations on 2020 August 23 and 24 UT and 2020 October 11 UT, TMO-LOT on 2020 September 22 UT, and MLO-GIT on 2020 October 16 and 20 UT.

\section{Results}
\subsection{Active and Inactive states of Gault}
\begin{figure}
    \centering
    \begin{tabular}{cc}
        \includegraphics[width=0.45\linewidth, height=5cm]{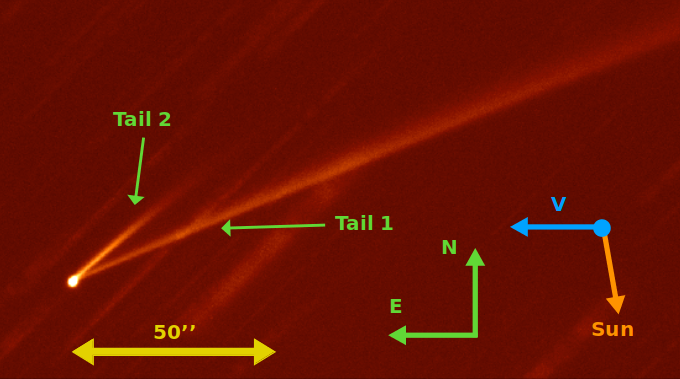} & \includegraphics[width=0.45\linewidth, height=5cm]{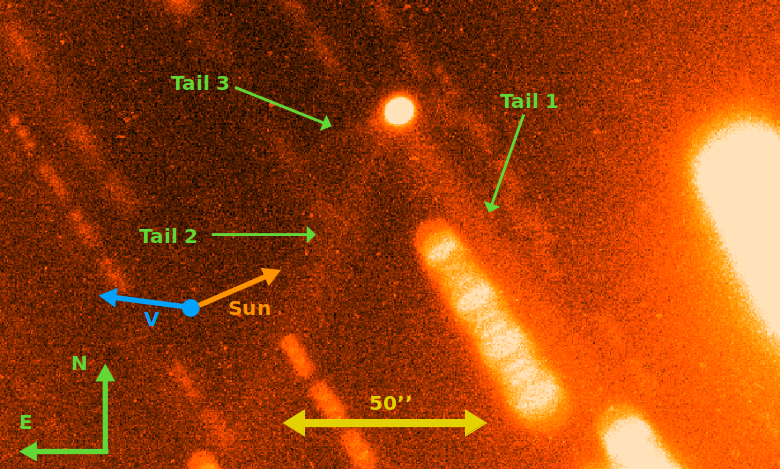} 
    \end{tabular}
    \caption{\textit{Left}: 4680-second Deep-stack image of Gault from the ARC 3.5-meter telescope at Apache Point Observatory on 2019 February 25 UT. The two tails indicate two separate epochs of activity. \textit{Right}: 4320-second Deep-stack image of Gault from the ARC 3.5-meter on 2019 April 26 UT after it had produced a third tail.}
    \label{Active}
\end{figure}

Follow-up observations with the ARC 3.5-meter telescope at Apache Point Observatory on 2019 February 25 UT showed evidence for multiple tails. A deep-stack image consisting of 4680 seconds of exposure is shown on the left side of Figure \ref{Active}. The surface brightness of Gault in this stack is 24.0 mag/arcsec$^2$ within a 10,000 km radius aperture. The right side of Figure \ref{Active} shows Gault's third tail in a 4320-second deep-stack image from the ARC 3.5-meter telescope on 2019 April 26 UT and the 10,000 km surface brightness was calculated to be 23.8 mag/arcsec$^2$.
The images were combined in deep, median-stacks centered on Gault and then used to compute the calibrated $r$-band photometry from comparisons to a similar deep-stacked images of reference stars with Solar colors from the same initial image. We referenced photometry of the reference star from the Pan-STARRS catalog \citep{Chambers_2016}.

\begin{figure}
    \centering
    \includegraphics[width=0.5\linewidth]{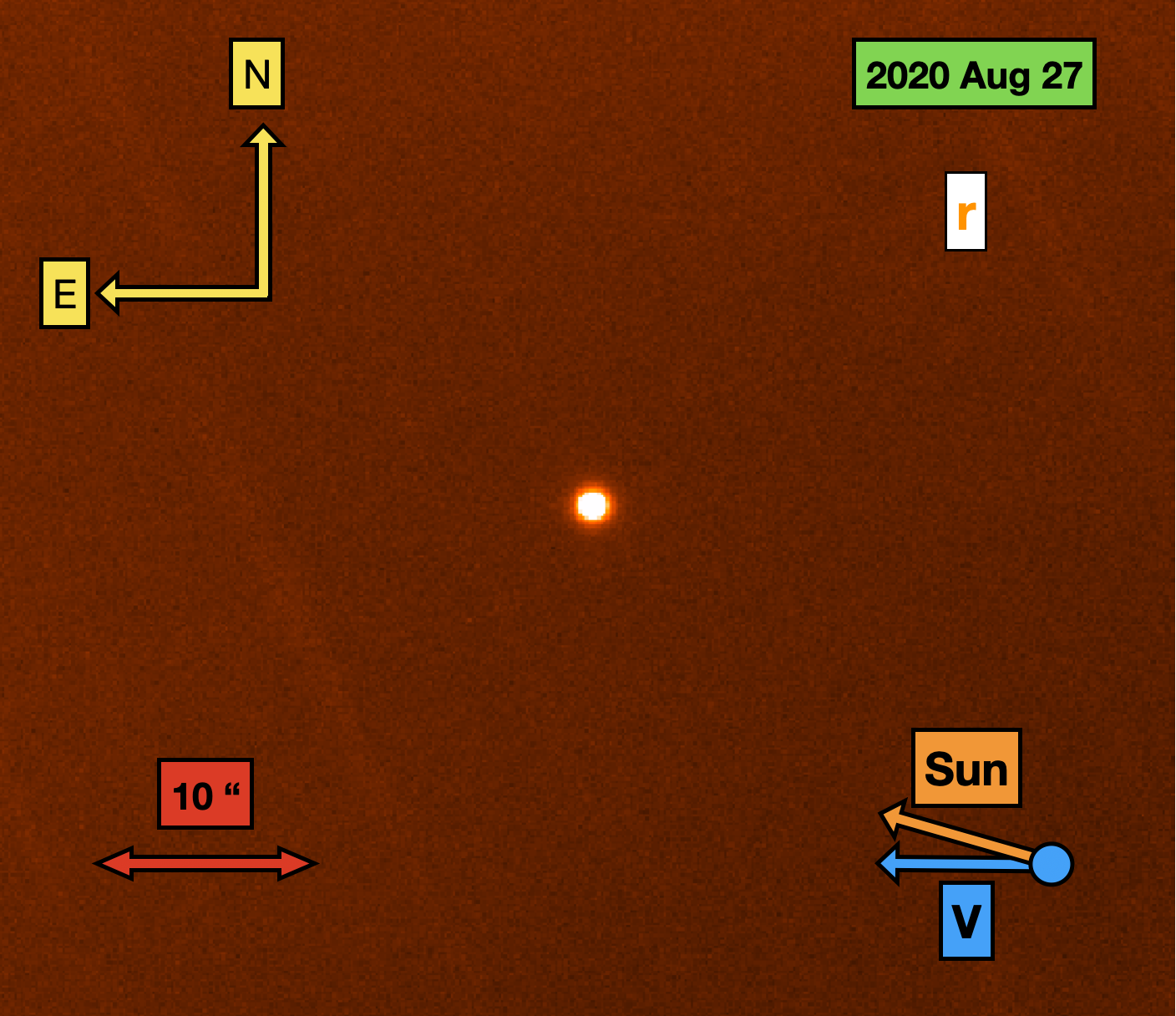}
    \caption{Deep-stack image of Gault taken with the P200 telescope in 103 90-second images culminating in 9270-seconds on 2020 August 27 UT in the \textit{r} band. The image displays a lack of coma or southwest-facing tail implying the inactivity of Gault.}
    \label{P200}
\end{figure}

Both surface brightnesses were brighter than the surface brightness of 25.8 mag/arcsec$^2$ found when it was inactive in 2020 June by \cite{Purdum_2020}. Our deep-stack image taken by the P200 in the \textit{r} band is shown in Figure \ref{P200}, which lacks cometary features and has a surface brightness of 26.3 mag/arcsec$^2$, also dimmer than the surface brightnesses from early 2019.

\subsection{Secular Photometry and Updated Absolute Magnitude}

\begin{figure}
    \centering
    \includegraphics[width=.7\linewidth]{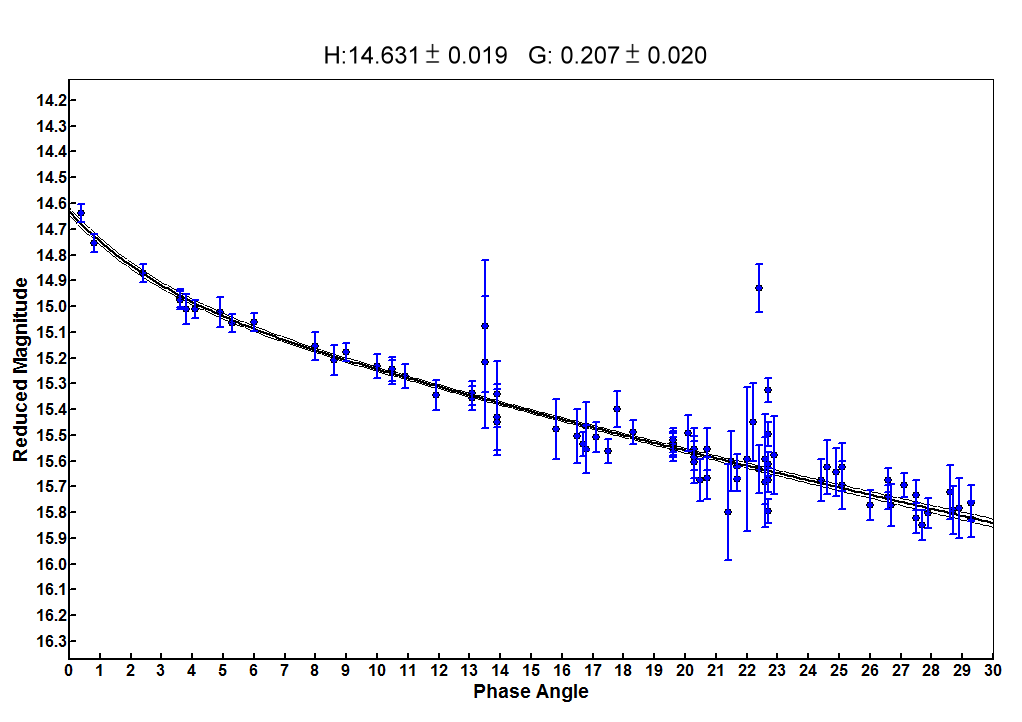}
    \caption{Reduced magnitude of Gault from Equation \ref{Eq_H} as a function of phase angle. The data points are taken from P48 observations with ZTF starting 2020 April 02 UT and ending 2020 October 14 UT. The line of best fit is based on $r - 5\, \mathrm{log_{10}}(R \Delta)$ in Equation \ref{Eq_H} using $H_r = 14.631 \pm 0.019$ and $G=0.207 \pm 0.020$.}
    \label{Phase_HG}
\end{figure}

Figure \ref{Phase_HG} shows Gault's reduced magnitude phasecurve from ZTF data taken between 2020 April 02 UT to 2020 October 14 UT. The reduced magnitude data is described by $r - 5\, \mathrm{log_{10}}(R \Delta)$ where $R$ and $\Delta$ are the heliocentric and geocentric distances. We can then find the $r$-band absolute magnitude $H_r$ and phase slope parameter $G$ by fitting the reduced magnitudes to the phase function of the form:

\begin{equation}
    r - 5\, \mathrm{log_{10}}(R \Delta)  = H - 2.5\,\mathrm{log_{10}}\left[ (1 - G)\,\Phi_1(\alpha) + G\,\Phi_2(\alpha) \right ]
    \label{Eq_H}  
\end{equation}
where $r$ is the $r$-band magnitude of Gault taken with a 5\arcsec~radius aperture, $\alpha$ is the phase angle of the asteroid at the time of mid-exposure, and  $\Phi_1$ and $\Phi_2$ are two basis functions normalized at unity for $\alpha = 0^{\circ}$ \citep[][]{Bowell_1989,Muinonen_2010,Pravec_2012}. 

\begin{figure}
    \centering
    \includegraphics[width=\linewidth]{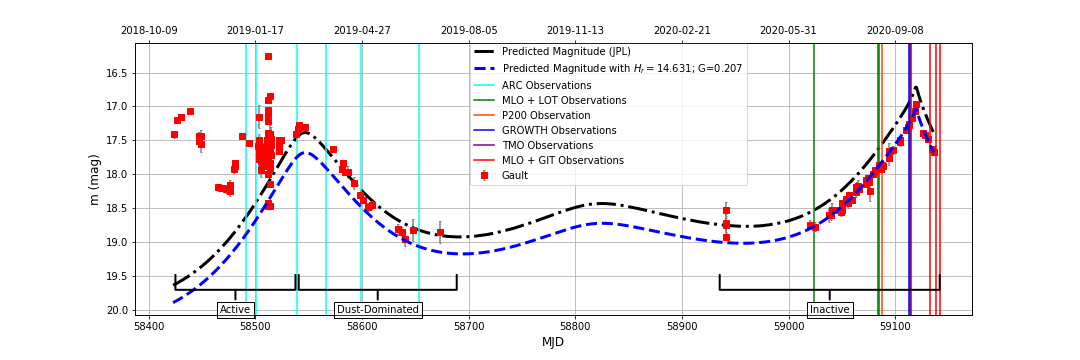}
    \caption{Secular photometry of Gault from 2018 November 01 UT (MJD = 58423) to 2020 October 14 UT (MJD = 59136). The red points are the apparent magnitudes of Gault taken by the Palomar Observatory 48-inch telescope and ZTF in the \textit{r} band over this time span. Data presented from before 2019 February 10 UTC are adapted from \citet[][]{Ye_2019}. The black line is the predicted apparent V-band magnitude from JPL HORIZON's ephemeris service. The blue line indicates what the predicted magnitude should be given new values for the absolute magnitude $H=14.631 \pm 0.019$ and slope parameter $G=0.207 \pm 0.020$. While the asteroid is active, the data do not line up with the predicted magnitude due to dust obscuring the surface of the asteroid. As Gault continued to be dust dominated as it entered Solar conjunction before MJD 58700, the predicted and observed magnitudes started to align again. After exiting Solar conjunction, Gault shows that it aligns with the new predicted values during inactivity. The vertical lines indicate the observation epochs in this work. }
    \label{ZTF_Phase}
\end{figure}

These best fit parameters are $H_r$ = 14.631 $\pm$ 0.019 and $G_r$ = 0.207 $\pm$ 0.020 after the asteroid had exited Solar conjunction and was no longer active. This is significantly fainter than the value of 14.31 $\pm$ 0.01 measured by ZTF when Gault was last seen to be inactive in 2017 \citep{Ye_2019}. This could be caused by the change in observing geometry over time creating a different line-of-site projection of Gault's light-scattering cross-section. The lack of a large lightcurve amplitude for Gault discussed below implies that the detection of Gault at the limiting magnitude of survey at the limits of the phasecurve was due to the viewing geometry of Gault rather than rotatioanal variations in its brightness \citep[][]{Jedicke2016}. Using the updated predicted absolute magnitude of Gault with the blue line in Figure \ref{ZTF_Phase}, we can see the instability in the brightness of Gault over time using 5\arcsec~radius aperture photometry. 

Figure \ref{ZTF_Phase} shows the time-series \textit{r} band ZTF photometry between June and October of 2020.
Gault exited Solar conjunction and was observed between 2020 April 02 UT (MJD = 58941) and 2020 October 14 UT (MJD = 59136). Comparing the measured equivalent r-band magnitude of Gault from photometry measured in ZTF observations taken on these dates with the predicted magnitude of Gault based on our measured r-band absolute magnitude of $H_r = 14.631 \pm 0.019$ and phase function slope value $G = 0.207 \pm 0.020$, we do not see any significant brightening in the actual magnitude of Gault compared to the predicted magnitudes. This is in contrast to the brightness of Gault in the ``active" portion of its lightcurve between 2018 November 01 UT (MJD = 58423) and 2019 February 10 UT (MJD = 58524). Between 2019 February 24 UT (MJD = 58538) and 2019 July 9 UT (MJD = 58673), the measured brightness and predicted brightness begin to become similar suggesting that the enhanced cross-section of Gault caused by dust within its vicinity was beginning to diminish in contrast to the increase in brightness of comets whose brightness are observed to increase with steadily increasing activity as they approach the Sun \citep[][]{Bolin2020asdfasdf,Bolin2021}.

\subsection{Time-Series Lightcurves}
\begin{figure}
    \centering
    \includegraphics[width=\linewidth]{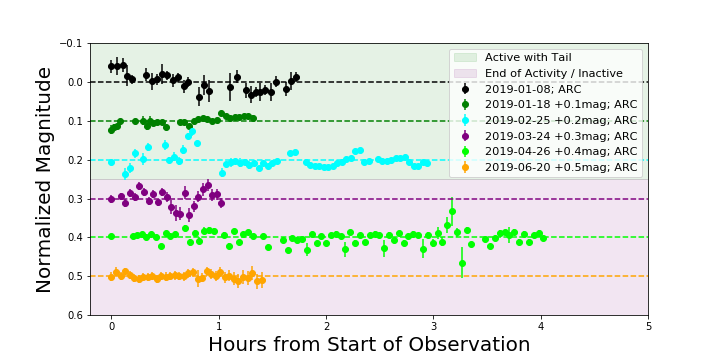}
    \caption{2019 data from the ARC 3.5-meter telescope. The data are organized chronologically from top to bottom and are offset by 0.1 magnitude from each other. The first 3 dates correspond to the epochs in which Gault was active and where the ZTF photometry is brighter than predicted, and the last 3 dates were obtained during the dust-dominated epochs that followed, as seen in Figure \ref{ZTF_Phase}.}
    \label{ARC_lightcurve}
\end{figure}

While Gault was active in 2019, the ARC 3.5-meter telescope at Apache Point Observatory took short-period lightcurve images, as shown in Figure \ref{ARC_lightcurve}. Much like Figure 6 in \cite{Jewitt_2019b}, this plot shows little variation in the lightcurve while the asteroid was experiencing activity and producing comet-like features. The variations that do occur in these lightcurves have small-amplitude peaks and are caused by noise consistent with the uncertainty values in their individual differential photometry. 

\begin{figure}
    \centering
    \includegraphics[width=\linewidth, height=21cm]{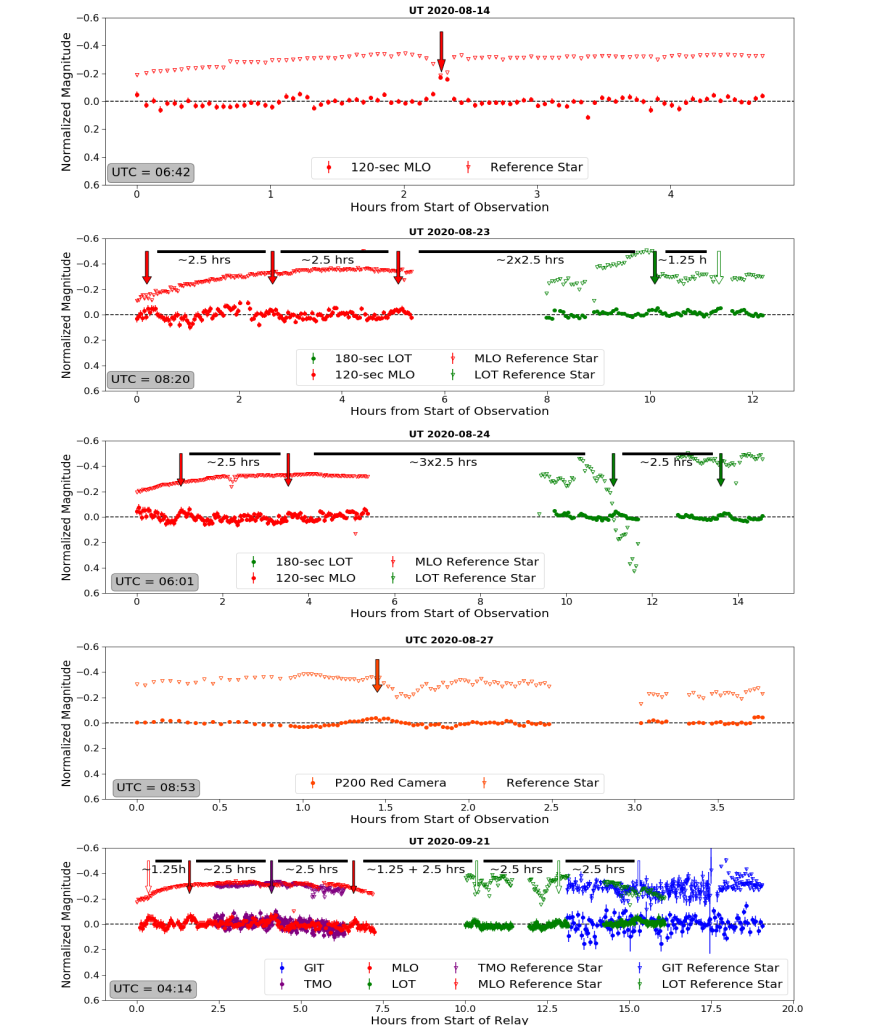}
    \caption{Coordinated long-term photometric lightcurves of Gault and various reference stars (offset -0.3mag) starting from 2020 August 14 UT and organized chronologically to 2020 September 21 UT. The observatories are color-coordinated and the bumps in the lightcurve caused by the rotation of $\sim2.5$ hours is denoted with arrows. The primary peaks are marked by solid arrows while secondary half-period peaks are denoted with white arrows. The lightcurves continue through 2020 October 20 UT in Figure \ref{Gault_lightcurves2}.}
    \label{Gault_lightcurves}
\end{figure}

\begin{figure}
    \centering
    \includegraphics[width=\linewidth, height=21cm]{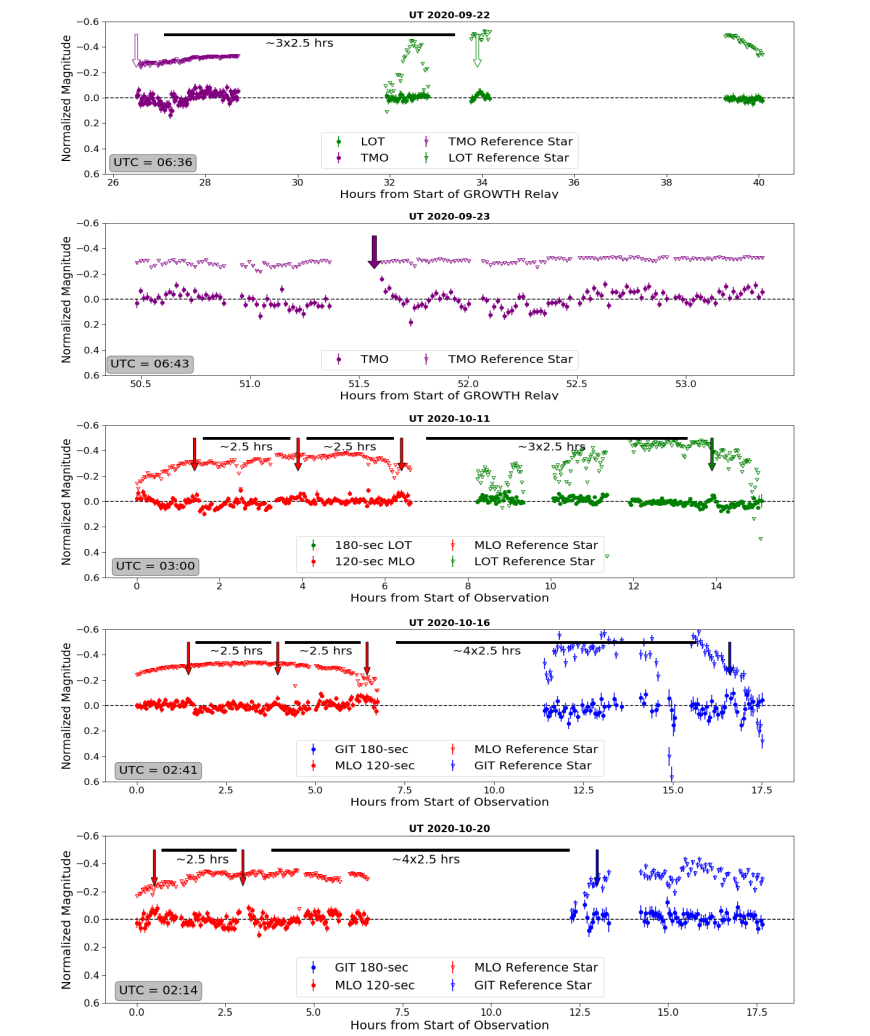}
    \caption{Coordinated long-term photometric lightcurves of Gault and various reference stars (offset -0.3mag) starting from 2020 September 22 UT and organized chronologically to 2020 October 20 UT. This figure is a continuation of Figure \ref{Gault_lightcurves} where the observatories are color-coordinated and the bumps in the lightcurve caused by the rotation of $\sim2.5$ hours is denoted with arrows. The primary peaks are marked by solid arrows while secondary half-period peaks are denoted with white arrows.}
    \label{Gault_lightcurves2}
\end{figure}

The flatness of the lightcurves is noted even as the brightness of Gault began to resemble its predicted brightness based on its pre-activity $H_r$ seen in Figure \ref{ZTF_Phase}. In order to determine the rotational period of Gault after it returned to an inactive state, we obtained coordinated long-term lightcurves on nine separate dates starting in August 2020, with the longest single lightcurve of 19 hours on 2020 September 21 UT. The results are displayed in Figures \ref{Gault_lightcurves} and \ref{Gault_lightcurves2}. 

The separate observatories are indicated by color, with MLO as red, LOT as green, P200 as orange, TMO as purple, and GIT as blue. Each lightcurve for each observatory has an additional reference star lightcurve normalized to -0.3 magnitude to show that the reference stars used to calculate the differential photometry of Gault did not vary over time (or if so, not at the same period and amplitude as the asteroid's lightcurves).  The reference star lightcurves have small error bars and have not been corrected for airmass or weather effects on their photometry and therefore are curved, but display no strong signs of periodicity. 

\begin{figure}
    \centering
    \includegraphics[width=\linewidth]{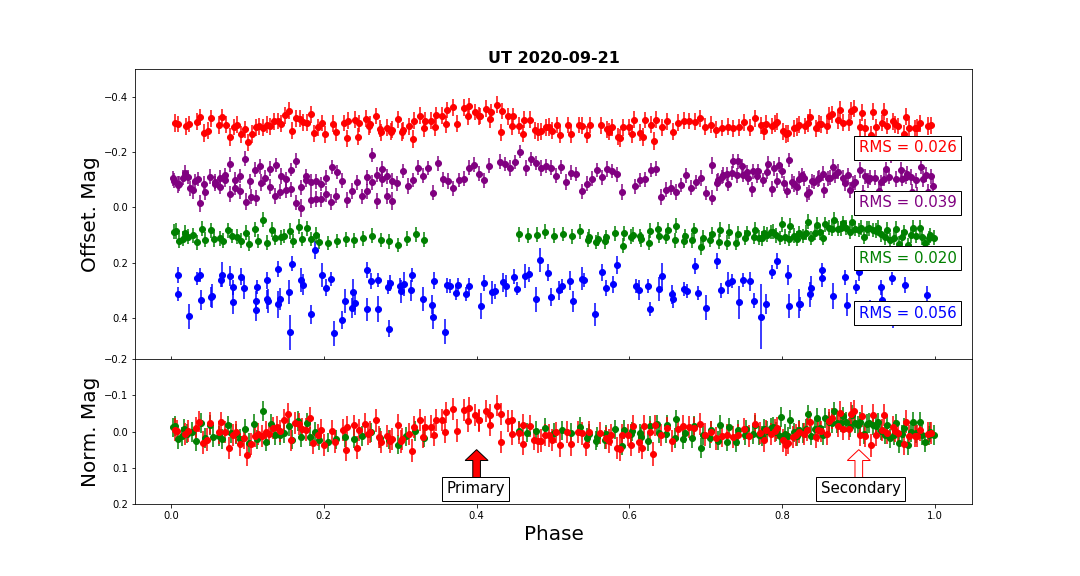}
    \caption{Phase-folded lightcurve of Gault on 2020 September 21 UT from the GROWTH relay of observations. Each telescope's lightcurve was folded by 2.5 hours and the MLO and LOT data were stacked to display the primary and secondary peaks of the double-peaked lightcurve. The root-mean-squared value is shown for each lightcurve as a reference for the amplitude of the noise in the data.}
    \label{GROWTH_Phase}
\end{figure}

Each date in Figures \ref{Gault_lightcurves} and \ref{Gault_lightcurves2} also arrows to indicate the small-amplitude, periodic peaks in the asteroid lightcurves based on multiples of 2.5-hour intervals from each other. Solid-colored arrows denote the primary peaks that represent a single rotation of the asteroid, while white arrows point to the half-period peaks at multiples of 1.25 hours from the primary peaks. Due to the minuscule ~0.1-magnitude amplitude of the peaks, many of them are overcome by the noise in the photometry and therefore only a handful appear on each date. Figure \ref{GROWTH_Phase} shows the an example of phase-folded lightcurves from the GROWTH relay of observations on 2020 September 21 UT. Like Figures \ref{Gault_lightcurves} and \ref{Gault_lightcurves2}, the telescopes are color coordinated and their lightcurves are folded by the double-peaked 2.5 hour period. The MLO and LOT data are stacked in the bottom plot and display modest primary and secondary peaks caused by Gault's periodic rotation. Additional phase-folded lightcurves of each observation date in Figures \ref{Gault_lightcurves} and \ref{Gault_lightcurves2} can be found in Figures \ref{0814_Phase} - \ref{1020_Phase} in the Appendix.

\begin{figure} 
    \centering
    \begin{tabular}{cc}
        \includegraphics[width=0.5\linewidth, height=5cm]{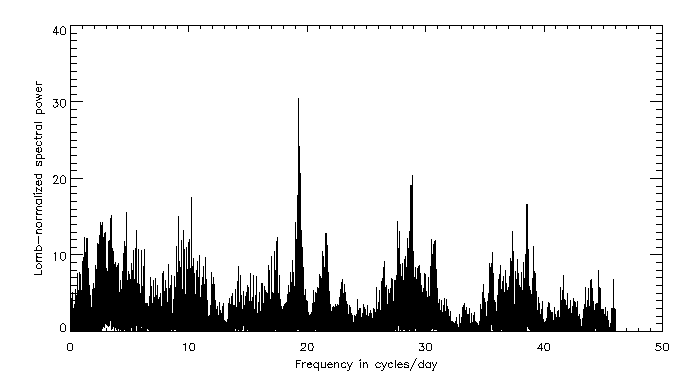} & \includegraphics[width=0.5\linewidth, height=5cm]{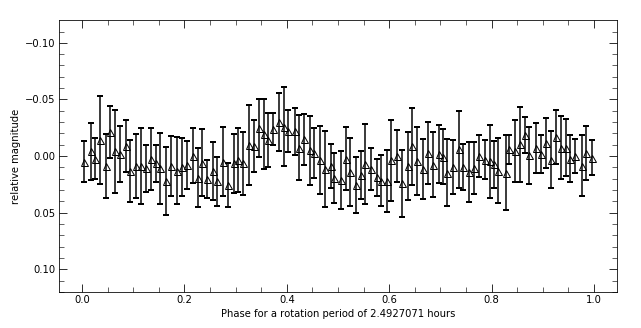}
    \end{tabular}
    
    \caption{\textit{Left}: Lomb-Scargle Periodogram of Gault's lightcurve data from MLO and LOT observations starting 2020 August 23 UT and ending 2020 October 20 UT. \textit{Right}: Folded lightcurve of Gault from MLO and LOT with a period of 2.5 hours averaged over 10 invidual MLO and LOT ligthcurves taken over six nights of data taken between 2020 August 23 UT and ending 2020 October 20 UT.}
    \label{LombScarg}
\end{figure}

A Lomb-Scargle \citep{Lomb_1976} periodogram was constructed from MLO and LOT data of Gault ranging from August to October of 2020 and is shown along with a folded lightcurve of Gault in Figure \ref{LombScarg}. The differential photometry technique we used to create the periodogram relied on determining the difference between the brightness of Gault and the comparison stars in the same field of view to acquire Gault’s light curve. The comparison stars we selected depended on the maximum frame width that we used through one run and on the similarity of FWHM (Full Width at Half Maximum) estimated between Gault and the chosen stars. By comparing the reference stars’ light curves, some variable stars were ruled out in the photometric analysis. To combine several photometry runs through different nights, the mean values of each run were automatically scaled using an IDL routine that we created. We then searched for significant periodicities using the Lomb-Scargle periodogram functions on the combined light curve data to find the most likely rotation period of Gault. The frequency analysis from the strong peak near $\sim20$ cycles/day in the left panel of Figure \ref{LombScarg} gives a rotation period of $\sim$1.25 hours, which corresponds to a single-peaked lightcurve. It is natural to assume a double-peaked lightcurve for Gault, which computes to a sidereal rotation period of 2.49 $\pm$ 0.07 hours. The right panel of Figure \ref{LombScarg} displays the lightcurve folded by the rotation period of 2.49 hours, which corresponds to a double-peaked lightcurve. The uncertainty here of $\pm 0.07$ hours is estimated using the bootstrap method \citep{Press_1986} which removed $\sqrt{N}$ data points from the time series lightcurves and recalculated the period value from the Lomb-Scargle periodogram. This process was repeated 10,000 separate times with the resulting central value of 2.49 hours and a $1-\sigma$ uncertainty estimate of 0.07 hours.

The amplitude of the MLO and LOT lightcurves are low, so it is somewhat difficult to recognize a continuous variation in brightness in the folded version of the lightcurve, as seen on the right in Figure \ref{LombScarg}. Here, the folded phase curves averaged by rebinnning them in phase space with a bin size of 0.01 and coadded into an average and the error bars are the 1 sigma scatter in data per phase bin.

\section{Discussion and Summary}
The surface brightness profiles taken from deep-stacked images of Gault from the MLO 1.0-meter on 2020 July 21 UT \citep{Purdum_2020} and P200 on 2020 August 27 UT indicate that Gault is no longer active after it appeared to have an outburst of material that caused multiple tails to form starting in October of 2018 \citep{Ye_2019}. The surface brightness measurements of 25.8 mag/arcsec$^2$ and 26.3 mag/arcsec$^2$, from MLO and Palomar, respectively, are fainter compared to the surface brightness values of Gault from the ARC 3.5-meter Telescope on 2019 February 9 UT, had a measurement of 24 mag/arcsec$^2$ \citep{Purdum_2020}. The fainter measurements in 2020 could mean that Gault no longer has material surrounding it and can be deemed inactive. 

Gault's deactivation can also be seen over time in the  photometry from ZTF observations of Gault in Figure \ref{ZTF_Phase}. The activity of Gault is apparent on the left side of the plot with the data being much brighter and more variable than the predicted magnitudes from JPL's HORIZONS ephemeris service\footnote{https://ssd.jpl.nasa.gov/?horizons}. The outbursts for Gault's first two tails were estimated to have occurred on 2018 October 18 $\pm$ 5 UT and 2018 December 24 $\pm$ 1 UT \citep{Ye_2019}, which is during the ``active" portion of the lightcurve (see Figure \ref{ZTF_Phase}). The third tail was much dimmer than the first two and so the exact time of initiation is uncertain, but \cite{Jewitt_2019b} estimates it to be 2019 February 10 $\pm$ 7 UT, right at the time the ``active" portion of Figure \ref{ZTF_Phase} ends. 

After Gault had produced its third tail, the ZTF photometry started to line back up with the predicted magnitudes from JPL's HORIZONS ephemeris service, indicating that the active stage had come to an end. However, this alignment did not last as the photometry from ZTF started to dip below the predicted magnitudes in Figure \ref{ZTF_Phase}. The misalignment occurred while Gault no longer had tails but was still surrounded by dust, which can skew photometric measurements. The right side of Figure \ref{ZTF_Phase}, however, shows that Gault's photometry exiting Solar conjunction was more stable than the active stage of the plot, therefore providing more evidence for Gault's inactivity. 

Interestingly, Gault no longer aligned with the predicted magnitudes HORIZONS ephemeris service after it had exited Solar conjunction. We found that altering the phase parameter $G$ in Equation \ref{Eq_H} from 0.25 to 0.21 and the absolute magnitude $H$ from JPL's HORIZONS' 14.3 to 14.6, re-aligns the photometry in Figure \ref{ZTF_Phase}. This phenomenon can be attributed to the observing geometry of Gault changing throughout its orbit. The brighter absolute magnitude $H_r$ during the ARC observations in 2019 (see Figure \ref{ZTF_Phase}) could be due to pole-on observations, while observations during different viewing geometries would result in smaller absolute magnitudes in 2020. The JPL HORIZONS' ephemeris service shows the estimated ecliptic longitude of Gault during observations by \cite{Ye_2019} in November of 2017 were $\sim 58\deg$ while our observations range from $\sim 3\deg$ -- $\sim 5 \deg$.

Photometric lightcurve observations with the ARC 3.5-meter while Gault was still dust-dominated (see Figure \ref{ARC_lightcurve}) in 2019 started to show some variation as the activity on Gault diminished, but the low-amplitude of the variations were not enough to conclude a rotation period. Our observations of Gault in 2020 also produced low-amplitude lightcurves (see Figures \ref{Gault_lightcurves} and \ref{Gault_lightcurves2}), even though it no longer displayed signs of activity. The viewing geometry is also much different for our observations in 2020 than from when it was active in 2019. This means that Gault could have a spherical or top-shaped geometry like Near-Earth Asteroids Ryugu and Bennu \citep{Hirabayashi_2020}. It is also worth noting that near-Earth Asteroid (3200) Phaethon was imaged by Arecibo radar observations and was found to have a round, top-like shape when it passed by Earth in December of 2017 \citep{Taylor_2019}. At that time, \cite{Kim_2018} found Phaethon's peak amplitude was small, $\sim$0.1 magnitudes, which is somewhat similar to our data in Figures \ref{Gault_lightcurves} and \ref{Gault_lightcurves2}.

Despite the low amplitude in the lightcurves, we did notice periodic small-amplitude peaks and found they are separated by roughly 2.5 hours and placed arrows in Figures \ref{Gault_lightcurves} and \ref{Gault_lightcurves2} to indicate the estimations. Some secondary peaks occurred at 1.25 hour intervals from the primary peaks due to the asteroid's geometry displaying a double-peaked lightcurve. Several peaks in the Gault lightcurves show little to no variation in the reference star lightcurve (see August 24, September 21, September 23, October 20 in Figures \ref{Gault_lightcurves} and \ref{Gault_lightcurves2}), which is an indication that the peaks are non-anomalous and are a result of the rotation of the asteroid. The periodogram in Figure \ref{LombScarg} shows Gault has a rotation period of 2.49 $\pm$ 0.07 hours, assuming the frequency of $\sim20$ cycles/day ($\sim$1.25 hours) corresponds to a double-peaked lightcurve. Figure \ref{GROWTH_Phase} and Figures \ref{0814_Phase} - \ref{1020_Phase} in the Appendix show the lightcurves for our observations spanning 2020 August 14 UT to 2020 October 20 UT (see Figures \ref{Gault_lightcurves} and \ref{Gault_lightcurves2}) folded by a period of 2.5 hours. The stronger primary peak and fainter secondary peak of Gault's double-peak lightcurve are denoted with arrows and the each Figure shows the lightcurves stacked by date. A root-mean-square calculation was made for each phase-folded lightcurve as a reference for the noise amplitude when considering looking at the primary and secondary bumps. Some dates show the peaks clearly, while others do not due to observation quality. Since the single-peak period of 1.25-hours from our Lomb-Scargle periodogram is aphysical for an asteroid of Gault's geometry, we only display the double-peak phase folded lightcurves.

\begin{figure}
    \centering
    \includegraphics[width=0.75\linewidth]{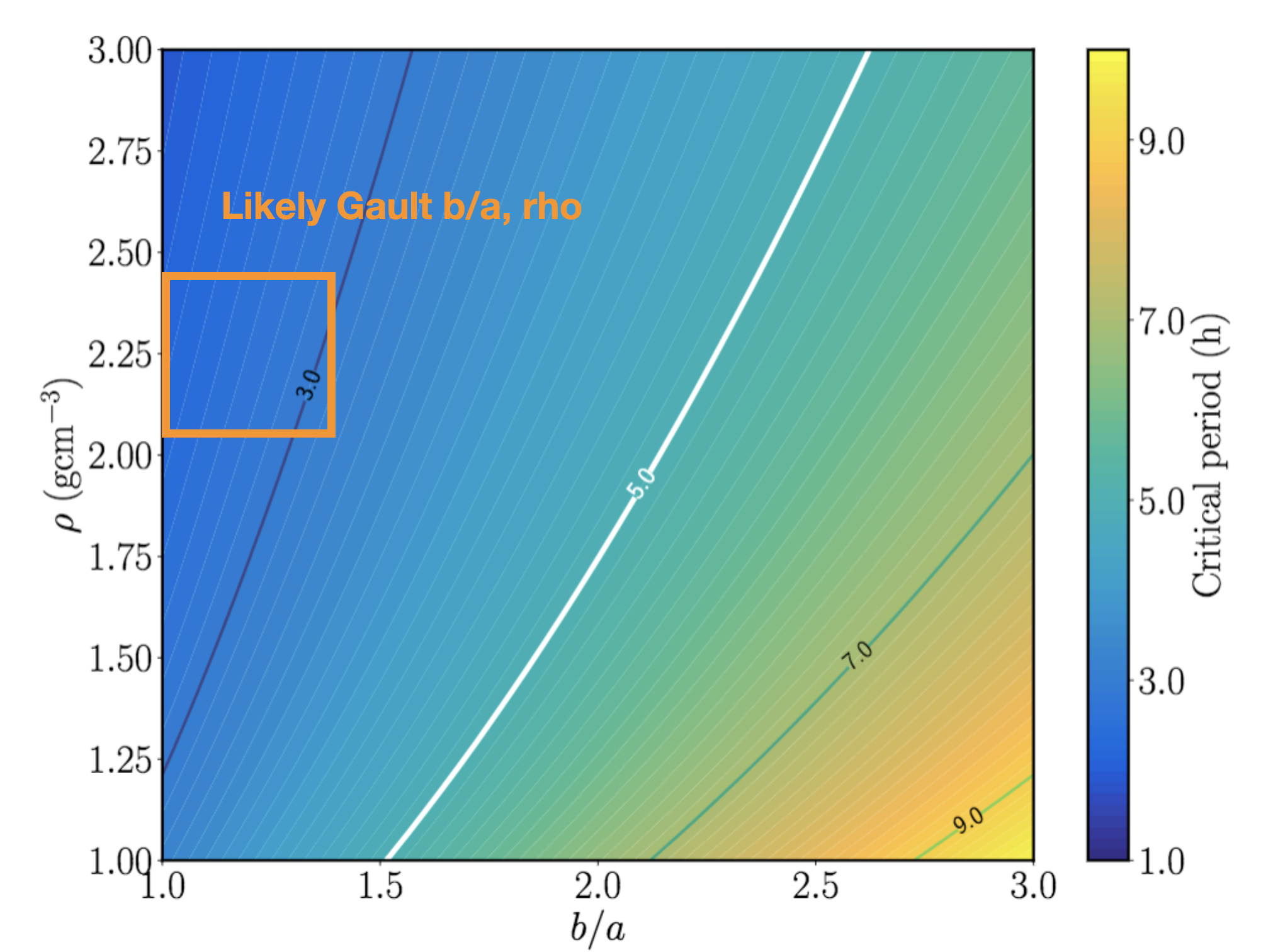}
    \caption{Adapted from \citet[][]{Bolin2018}. The critical rotation period of an asteroid based on the axial ratio b/a and the density $\rho$. The likely critical period for Gault is indicated in the orange box, roughly around 2.0 - 3.0 hours.}
    \label{Crit_period}
\end{figure}

With the assumption that Gault is nearly spherical or top-shaped (see \cite{Harris_2014} on the determination of an asteroid's shape from its lightcurve), we assume the $b$/$a$ axial ratio is close to 1--1.3 and $b$/$c$ $\sim $1.3 given its maximum possible lightcurve amplitude of $\sim$0.1-0.3 as seen in our data and the data from \citet[][]{Kleyna_2019}\footnote{Although \cite{Kleyna_2019} show a similar small-amplitude magnitude variation to our results, it is worth noting that their results were found while Gault was showing signs of activity.} and the relation between $b$/$a$ and lightcurve amplitude of $b$/$a$ = 10$^{0.4\Delta M}$ from \citet[][]{Binzel1989}. It should be noted that the axial ratio inferred from the observed lightcurve amplitude can be affected by the angle between the spin pole and the observer \citep[][]{Vokrouhlicky2017a, Hanus2018}, however, the consistently small lightcurve amplitude from different viewing geometries in the 2019 and 2020 apparitions seems to favor a smaller axial ratio. The critical breakup period of a strengthless ellipsoid as a function of axial ratio is given by \cite{Jewitt_2017} as 

\begin{equation}
    P_{crit} = \bigg( {b \over a} \bigg)\bigg[{{3\pi} \over {G\rho}} \bigg]^{1/2}
    \label{Eq_Period}
\end{equation}
where $\rho$ is the density of the ellipsoid and $G$ is the Newtonian gravitational constant. Gault should have a density of roughly equal to 2.2~g/cm$^3$ \citep{Sanchez_2019,Marsset2019}, consistent with other S-Type asteroids \citep{CARRY_2012}. Figure \ref{Crit_period} presents the critical period with which Gault would start shedding surface material as a function of the axial ratio and density. 
An orange box labelled in Figure \ref{Crit_period} is likely to contain the critical rotation period for an object with Gault's geometry. A rotation period like the one we have found at 2.5 hours seems to be at or near the critical period of Gault and therefore could be the cause of the activity started in 2018. Previous authors have also proposed Gault's activity was caused by rotational instability induced from the YORP effect \citep{Jewitt_2019b, Kleyna_2019, Ferrin_2019}. \cite{Jewitt_2015} determined that the YORP spin-up timescale for Gault should be roughly 22 Myr, much shorter than the 100 Myr-timescale for the re-orientation of the spin of a $\sim$4 km asteroid by non-destructive collisions \citep[][]{Farinella1998}. It is therefore possible that Gault may experience rotational fission if it continues to be spun up past its rotational breakup limit \citep[eg.,][]{Jewitt2017,Moreno2017}. \cite{Scheeres_2015} theorize that the surface of an asteroid after YORP spin-up can be ``perched" and ready for failure in multiple areas, which can cause multiple epochs of activity similar to active asteroid P/2013 P5 \citep{Jewitt_2013, Hainaut_2014}. We believe Gault could have had multiple epochs of activity after an initial YORP spin-up despite losing angular momentum from ejected particles.

Some asteroids are observed to be spinning beyond their rotational limit, but this is typically for asteroids smaller than the kilometer scale \citep[][]{Pravec2008} to smaller than the meter scale \citep[][]{Bolin2014,Bolin2020CD3} which are held together by cohesive forces \citep[][]{Sanchez2014}. However, it should be noted that examples of km-scale asteroids have been found rotating faster than their critical period \citep[][]{Chang2017} from fast rotating asteroid searches in wide-field optical surveys \citep[][]{Chang2016,Chang2019}. 

Other causes for activity are somewhat less likely than rotational instability. For instance, unlike Main Asteroid Belt Comets \citep{Jewitt_2015}, there are many clues that point away from ice sublimation as the driver for Gault's activity. Since Gault is an S-type and a member of the inner Main Belt Phocaea family, it likely formed inside the snow line and therefore would not experience sublimation \citep{Vernazza_2016}. Additionally, Gault's activity was observed with the ARC 3.5-meter telescope when Gault was located at a heliocentric distance of 2.41 AU on 2019 February 9 UT. When Gault was observed to be inactive by the MLO 1.0-meter telescope on 2020 June 24 UT, it had a heliocentric distance of 2.07 AU, suggesting the shorter distance to perihelion did not drive the activity. Collision events are also unlikely to be the cause for activity due to the multiple epochs of activity occurring in the few months Gault was active. Thermal disintegration is also somewhat unlikely due to Gault's low-eccentricity orbit. Active asteroids with this kind of driver typically have highly eccentric orbits which cause large changes in temperature that lead to fracture \citep{Delbo2014,Jewitt_2015}. Other processes such as Solar radiation pressure are unlikely due to Gault's rotation constantly changing the orientation with respect to the Sun. Therefore we find that Gault's activity was likely driven from rapid rotation at its critical period of 2.5 hours.

\acknowledgments
Based on observations obtained with the Samuel Oschin Telescope 48-inch and the 60-inch Telescope at the Palomar Observatory as part of the Zwicky Transient Facility project. ZTF is supported by the National Science Foundation under Grant No. AST-2034437 and a collaboration including Caltech, IPAC, the Weizmann Institute for Science, the Oskar Klein Center at Stockholm University, the University of Maryland, Deutsches Elektronen-Synchrotron and Humboldt University, the TANGO Consortium of Taiwan, the University of Wisconsin at Milwaukee, Trinity College Dublin, Lawrence Livermore National Laboratories, and IN2P3, France. Operations are conducted by COO, IPAC, and UW.

This work was supported by the GROWTH project funded by the National Science Foundation under PIRE Grant No 1545949.

Part of this work was performed under the auspices of the U.S. Department of Energy by Lawrence Livermore National Laboratory under Contract DE-AC52-07NA27344.

B.T.B. and F.J.M. acknowledge support from NASA with grant number 80NSSC19K0780.

C.F.~gratefully acknowledges the support of his research by the Heising-Simons Foundation ($\#$2018-0907).

M.~W.~Coughlin acknowledges support from the National Science Foundation with grant number PHY-2010970.

This publication has made use of data collected at Lulin Observatory, partly supported by MoST grant 108-2112-M-008-001.

C.C.N. thanks to the funding from MOST grant 104-2923-M-008-004-MY5.

C.Z. acknowledges support from JPL's internal research funds of the R$\&$TD, JROC, and ESD HBCU/MSI programs.

J. N. P. and R. Q. acknowledge support from JPL's ESD HBCU/MSI program under Subcontract 1659249.

The work of J.H. has been supported by the Czech Science Foundation through grant 20-08218S and by the Charles University Research program No. UNCE/SCI/023.

V.B., K.S., and H.K. thank Kunal Deshmukh for help with data processing. GROWTH India telescope is a 70-cm telescope with a 0.7 degree field of view, set up by the Indian Institute of Astrophysics and the Indian Institute of Technology Bombay with support from the Indo-US Science and Technology Forum (IUSSTF) and the Science and Engineering Research Board (SERB) of the Department of Science and Technology (DST), Government of India (https://sites.google.com/view/growthindia/). It is located at the Indian Astronomical Observatory (Hanle), operated by the Indian Institute of Astrophysics (IIA). GROWTH-India project is supported by SERB and administered by IUSSTF.

H.K. thanks the LSSTC Data Science Fellowship Program, which is funded by LSSTC, NSF Cybertraining Grant \#1829740, the Brinson Foundation, and the Moore Foundation; his participation in the program has benefited this work.

\facility{Apache Point Astrophysical Research Consortium 3.5 m telescope, GROWTH India Telescope, Lulin Optical Telescope, Mount Laguna Observatory 40-inch Telescope, P48 Oschin Schmidt telescope/Zwicky Transient Facility, Table Mountain Observatory}
\software{Astropy \citep{astropy:2013, astropy_2018}}, ZChecker \citep{Kelley_2019}, Aperture Photometry Tool \citep{Laher_2012}
\bibliography{bib}
\bibliographystyle{aasjournal}

\appendix

\begin{longtable}{|c|c|c|c|c|c|c|c|c|}
    \caption[width=\textwidth]{Telescope Specifications and Parameters for This Work} \\
    \hline
    \textbf{Telescope}$^1$ & \textbf{CCD}$^2$ & \textbf{Pixels}$^3$ & \textbf{Binning}$^4$ & \textbf{Scale (\arcsec/pix)}$^5$ & \textbf{Exp (s)}$^6$ & \textbf{NST}$^7$ \\
    \hline
    MLO 1.0-meter & ULTRAcam & 2Kx2K & 2x2 & 0.358 & 120 & N \\
    Lulin One-meter & SOPHIA & 2Kx2K & 1x1 & 0.385 & 180 & Y \\
    ARC 3.5-meter & ARCTIC & 2Kx2K & 2x2 & 0.228 & 120 & Y \\
    Palomar 200-inch & CHIMERA & 1Kx1K & 1x1 & 0.29 & 90 & Y \\
    Palomar 48-inch & ZTF CCD & 16 6Kx6K & 1x1 & 1.01 & 30 & N \\
    GIT 0.7-meter & Apogee KAF3200EB & 2Kx1K & 1x1 & 0.3 & 120-180 & N \\
    TMO 1.0-meter & sCMOS & 1.6Kx1.6K & 1x1 & 0.225 & 60 & N \\
    \hline
    \caption{Columns: (1) Telescope name; (2) CCD Camera name; (3) Number of Pixels on CCD; (4) Type of binning used; (5) Pixel scale; (6) Exposure Time; (7) Non-sidereal tracking enabled (Y/N).}
     \label{table1}
\end{longtable}

\setlength{\LTcapwidth}{6in}
\begin{longtable}{|c|c|c|c|c|c|c|c|c|}
    \caption[width=\textwidth]{Observations of Gault producing photometric lightcurves.} \\
    \hline
    Date (UTC)$^1$ & Telescope$^2$ & RA$^3$ & DEC$^4$ & $r$ (AU)$^5$ & $\Delta$ (AU)$^6$ & $\alpha$ ($^{\circ}$)$^7$ & filter$^8$ & $\theta_s$ (")$^9$\\
    \hline
         2019 Jan 08 & ARC & 10 48 15.82 & -12 34 36.1 & 2.470 & 1.865 & 20.8 & $r'$ & 1.3 \\ 
         
         2019 Jan 18 & ARC & 10 48 03.15 & -12 42 42.3 & 2.451 & 1.735 & 18.8 & $r'$ & 1.7\\
         
         2019 Feb 25 & ARC & 10 25 39.71 & -08 20 25.9 & 2.374 & 1.413 & 7.3 &  $r'$ & 1.4\\
         
         2019 Mar 24 & ARC & 10 04 25.84 & -01 08 19.5 & 2.316 & 1.406 & 12.9 & $r'$ & 1.2\\
         
         2019 Apr 26 & ARC & 10 02 29.07 & +06 19 04.2 & 2.243 & 1.619 & 23.9 & $r'$ & 2.5 \\
         
         2019 Jun 20 & ARC & 10 59 34.47 & +09 20 53.9 & 2.121 & 2.150 & 27.5 & $r'$ & 2.7\\
         
         2020 Jun 24 & MLO & 00 06 26.41 & +13 32 09.8 & 2.071 & 1.873 & 29.3 & $R$ & 2.2 \\
         
         2020 Aug 14 & MLO & 00 41 35.22 & +12 07 57.8 & 2.182 & 1.416 & 21.8 & $R$ & 1.7 \\
         
         2020 Aug 23 & MLO & 00 40 34.98 & +10 35 29.9 & 2.202 & 1.355 & 18.4 & $R$ & 2.2 \\
         
         2020 Aug 23 & LOT & 00 40 31.32 & +10 32 32.2 & 2.202 & 1.354 & 18.3 & $R$ & 2.8\\
         
         2020 Aug 24 & MLO & 00 40 19.44 & +10 23 30.5 & 2.204 & 1.349 & 18.0 & $R$ &  1.6\\
         
         2020 Aug 24 & LOT & 00 40 12.37 & +10 18 26.4 & 2.205 & 1.347 & 17.8 & $R$ & 2.5 \\
         
         2020 Aug 27 & P200 & 00 39 30.16 & +09 50 27.8 & 2.210 & 1.334 & 16.9 & $r$ &  1.1\\
         
         2020 Sep 21 & MLO & 00 24 06.34 & +03 10 17.1 & 2.265 & 1.268 & 4.0 & $R$ & 2.1\\
         
         2020 Sep 21 & TMO & 00 24 02.45 & +03 08 46.4 & 2.265 & 1.268 & 3.9 & $R$ & 1.2\\
         
         2020 Sep 21 & LOT & 00 23 22.57 & +02 53 41.4 & 2.267 & 1.268 & 3.4 & $R$ & 1.7\\
         
         2020 Sep 21 & GIT & 00 23 05.87 & +02 47 37.3 & 2.268 & 1.269 & 3.2 & $r'$ & 2.5\\
         
         2020 Sep 22 & TMO & 00 23 12.10 & +02 49 52.8 & 2.268 & 1.269 & 3.3 & $R$ & 1.6\\
         
         2020 Sep 22 & LOT & 00 22 39.51 & +02 37 47.3 & 2.269 & 1.269 & 2.9 & $R$ & 1.7 \\
         
         2020 Sep 23 & TMO & 00 22 23.43 & +02 31 43.6 & 2.270 & 1.270 & 2.7 & $R$ & 1.8\\
        
         2020 Sep 24 & TMO & 00 21 36.54 & +02 14 19.8 & 2.272 & 1.271 & 2.2 & $R$ & 1.2\\

         2020 Oct 11 & MLO & 00 08 16.22 & -02 44 28.5 & 2.310 & 1.339 & 7.5 & $R$ & 3.0 \\
         
         2020 Oct 11 & LOT & 00 08 02.98 & -02 49 37.2 & 2.311 & 1.341 & 7.6 & $R$ & 2.6\\
         
         2020 Oct 16 & MLO & 00 05 10.29 & -03 58 35.4 & 2.321 & 1.373 & 9.9 & $R$ & 2.0\\
         
         2020 Oct 16 & GIT & 00 04 45.91 & -04 08 29.4 & 2.323 & 1.379 & 10.3 & $r'$ & 3.1 \\
         
         2020 Oct 20 & MLO & 00 03 05.67 & -04 51 51.5 & 2.330 & 1.405 & 11.7 & $R$ & 2.6 \\
         
         2020 Oct 20 & GIT & 00 02 43.96 & -05 01 22.7 & 2.331 & 1.412 & 12.1 & $r'$ & 2.7\\
         \hline
    
    \caption{Columns: (1) Date of observation; (2) Telescope; (3) Right Ascension at the start of observation; (4) Declination at the start of observation; (5) Heliocentric distance at start of observation; (6) Geocentric distance at start of observation; (7) Phase Angle at start of observation; (8) Filter; (9) in-image seeing of at the start of observation.}
    \label{table2}
\end{longtable}

\begin{table}[h!]
    \centering
    \begin{tabular}{|c|c|c|c|}
    \hline
    \textbf{JD}$^1$ & \textbf{Mag}$^2$ & $\sigma_{\textbf{Mag}}$ $^3$ & \textbf{Observatory}$^4$ \\
    \hline
    2459075.77946 & 0.000828 & 0.025020 & MLO \\ 
    2459075.78247 & -0.012480 & 0.022338 & MLO \\ 
    2459075.78466 & 0.005480 & 0.023636 & MLO \\ 
    2459075.78683 & -0.001881 & 0.021923 & MLO \\ 
    2459075.78901 & 0.013656 & 0.023636 & MLO \\ 
    2459075.79117 & 0.002243 & 0.023137 & MLO \\ 
    2459075.79336 & 0.002003 & 0.022937 & MLO \\ 
    2459075.79554 & 0.007662 & 0.023037 & MLO \\ 
    2459075.79771 & -0.002169 & 0.021340 & MLO \\ 
    2459075.79988 & 0.007134 & 0.021839 & MLO \\ 
    \vdots & \vdots & \vdots & \vdots \\
    \hline
    \end{tabular}
    \caption{Photometric Lightcurve data for Figures \ref{Gault_lightcurves} and \ref{Gault_lightcurves2}. Columns: (1) Julian Date; (2) Normalized Magnitude; (3) Normalized Magnitude Uncertainty; (4) Location Observations were taken.}
    \tablecomments{Table 3 is published in its entirety in the machine-readable format.
    A portion is shown here for guidance regarding its form and content.}
    \label{table3}
\end{table}

\begin{longtable}{|c|c|c|c|c|c|c|c|c|}
    \caption{Observation data for the secular lightcurve taken between 2020 April 02 UT and 2020 October 14 UT by ZTF as shown in Figure \ref{ZTF_Phase}.} \\
    \hline
    \textbf{Date (UTC)}$^1$ & \textbf{RA}$^2$ & \textbf{DEC}$^3$ & $R$ (AU)$^4$ & $\Delta$ (AU)$^5$ & $\alpha$ ($^\circ$)$^6$ & \textbf{mag}$^7$ & $\chi_{am}^8$ & $\theta_s$ (\arcsec)$^9$\\
    \endfirsthead
    \caption* {Continued from last page} \\
    \hline
    \textbf{Date (UTC)}$^1$ & \textbf{RA}$^2$ & \textbf{DEC}$^3$ & $R$ (AU)$^4$ & $\Delta$ (AU)$^5$ & $\alpha$ ($^\circ$)$^6$ & \textbf{mag}$^7$ & $\chi_{am}^8$ & $\theta_s$ (\arcsec)$^9$\\
    \hline
    \endhead % all the lines above this will be repeated on every page
    \hline
    \multicolumn{5}{r@{}}{Continued on next page}\\
    \endfoot
    \endlastfoot
        \hline
        2020-04-02 12:14 & 21:43:56.3 & +02:13:35 & 1.924  & 2.480 & 21.9 & 18.72 $\pm$ 0.13   & 2.525 & 4.902 \\ 
        2020-04-02 12:21 & 21:43:56.9 & +02:13:38 & 1.924  & 2.480 & 21.9 & 18.74 $\pm$ 0.13   & 2.399 & 5.154 \\ 
        2020-04-02 12:27 & 21:43:57.4 & +02:13:41 & 1.924  & 2.480 & 21.9 & 18.53 $\pm$ 0.12   & 2.286 & 3.563 \\ 
        2020-04-02 12:34 & 21:43:58.0 & +02:13:43 & 1.924  & 2.480 & 21.9 & 18.93 $\pm$ 0.21   & 2.185 & 4.330 \\ 
        2020-06-24 11:24 & 00:06:30.7 & +13:32:24 & 2.071  & 1.872 & 29.3 & 18.77 $\pm$ 0.07  & 1.273 & 1.706 \\ 
        2020-07-08 09:03 & 00:21:56.5 & +14:10:18 & 2.101  & 1.742 & 28.8 & 18.60 $\pm$ 0.10   & 2.066 & 2.451 \\ 
        2020-07-09 10:56 & 00:22:59.0 & +14:11:41 & 2.103  & 1.732 & 28.7 & 18.60 $\pm$ 0.08   & 1.262 & 2.708 \\ 
        2020-07-10 09:17 & 00:23:52.1 & +14:12:40 & 2.105  & 1.723 & 28.6 & 18.52 $\pm$ 0.09  & 1.81  & 1.847 \\ 
        2020-07-17 10:05 & 00:29:55.6 & +14:13:52 & 2.120  & 1.657 & 27.9 & 18.53 $\pm$ 0.05   & 1.331 & 1.536 \\ 
        2020-07-19 09:20 & 00:31:25.3 & +14:12:06 & 2.124  & 1.639 & 27.7 & 18.56 $\pm$ 0.05  & 1.515 & 1.901 \\ 
        2020-07-20 11:26 & 00:32:12.3 & +14:10:43 & 2.127  & 1.628 & 27.5 & 18.43 $\pm$ 0.04  & 1.11  & 1.861 \\ 
        2020-07-20 11:30 & 00:32:12.4 & +14:10:43 & 2.127  & 1.628 & 27.5 & 18.52 $\pm$ 0.05  & 1.105 & 1.969 \\ 
        2020-07-23 11:03 & 00:34:12.6 & +14:05:21 & 2.133  & 1.601 & 27.1 & 18.36 $\pm$ 0.04  & 1.125 & 1.776 \\ 
        2020-07-25 11:56 & 00:35:26.9 & +14:00:20 & 2.138  & 1.582 & 26.7 & 18.42 $\pm$ 0.07  & 1.076 & 1.506 \\ 
        2020-07-26 11:06 & 00:36:00.0 & +13:57:33 & 2.140  & 1.574 & 26.6 & 18.38 $\pm$ 0.05   & 1.106 & 1.520 \\ 
        2020-07-26 11:35 & 00:36:00.6 & +13:57:29 & 2.140  & 1.573 & 26.6 & 18.31 $\pm$ 0.04   & 1.082 & 1.534 \\ 
        2020-07-29 11:02 & 00:37:33.1 & +13:47:16 & 2.146  & 1.547 & 26.0 & 18.38 $\pm$ 0.05  & 1.098 & 1.763 \\ 
        2020-08-02 08:51 & 00:39:12.9 & +13:29:57 & 2.155  & 1.513 & 25.1 & 18.26 $\pm$ 0.08  & 1.377 & 2.476 \\ 
        2020-08-02 09:58 & 00:39:13.8 & +13:29:43 & 2.155  & 1.512 & 25.1 & 18.19 $\pm$ 0.06  & 1.168 & 1.837 \\ 
        2020-08-03 08:59 & 00:39:34.4 & +13:24:46 & 2.157  & 1.504 & 24.9 & 18.20 $\pm$ 0.08   & 1.33  & 2.256 \\ 
        2020-08-04 09:58 & 00:39:54.8 & +13:19:04 & 2.160  & 1.495 & 24.6 & 18.17 $\pm$ 0.09   & 1.153 & 2.016 \\ 
        2020-08-05 11:04 & 00:40:13.4 & +13:13:01 & 2.162  & 1.486 & 24.4 & 18.21 $\pm$ 0.07  & 1.078 & 1.596 \\ 
        2020-08-11 11:31 & 00:41:24.0 & +12:31:26 & 2.175  & 1.438 & 22.7 & 18.15 $\pm$ 0.04  & 1.082 & 1.670 \\ 
        2020-08-11 11:46 & 00:41:24.0 & +12:31:21 & 2.175  & 1.437 & 22.7 & 18.09 $\pm$ 0.04  & 1.092 & 1.840 \\ 
        2020-08-12 11:34 & 00:41:29.5 & +12:23:22 & 2.178  & 1.430 & 22.4 & 18.10 $\pm$ 0.04    & 1.087 & 1.463 \\ 
        2020-08-13 10:19 & 00:41:33.2 & +12:15:26 & 2.180  & 1.423 & 22.0 & 18.05 $\pm$ 0.24   & 1.085 & 1.361 \\ 
        2020-08-14 11:06 & 00:41:35.1 & +12:06:27 & 2.182  & 1.415 & 21.7 & 18.12 $\pm$ 0.04  & 1.077 & 1.436 \\ 
        2020-08-14 11:39 & 00:41:35.1 & +12:06:15 & 2.182  & 1.415 & 21.7 & 18.07 $\pm$ 0.04  & 1.096 & 1.521 \\ 
        2020-08-15 09:49 & 00:41:35.4 & +11:57:55 & 2.184  & 1.408 & 21.4 & 18.24 $\pm$ 0.16  & 1.104 & 1.612 \\ 
        2020-08-18 09:37 & 00:41:25.2 & +11:28:55 & 2.191  & 1.387 & 20.3 & 18.00 $\pm$ 0.04   & 1.104 & 1.912 \\ 
        2020-08-20 09:38 & 00:41:09.4 & +11:07:50 & 2.195  & 1.373 & 19.6 & 17.95 $\pm$ 0.04  & 1.096 & 2.111 \\ 
        2020-08-20 10:05 & 00:41:09.2 & +11:07:38 & 2.195  & 1.373 & 19.6 & 17.93 $\pm$ 0.04  & 1.078 & 1.776 \\ 
        2020-08-23 11:39 & 00:40:31.0 & +10:32:42 & 2.202  & 1.354 & 18.3 & 17.86 $\pm$ 0.04  & 1.143 & 1.828 \\ 
        2020-08-25 08:49 & 00:39:59.9 & +10:09:44 & 2.206  & 1.342 & 17.5 & 17.92 $\pm$ 0.04  & 1.131 & 1.716 \\ 
        2020-08-26 09:03 & 00:39:40.5 & +09:56:56 & 2.209  & 1.337 & 17.1 & 17.86 $\pm$ 0.05  & 1.107 & 2.218 \\ 
        2020-08-27 09:29 & 00:39:19.3 & +09:43:41 & 2.211  & 1.331 & 16.7 & 17.88 $\pm$ 0.04  & 1.082 & 1.510 \\ 
        2020-09-02 09:11 & 00:36:41.5 & +08:19:24 & 2.224  & 1.303 & 13.9 & 17.76 $\pm$ 0.11  & 1.079 & 1.499 \\ 
        2020-09-02 09:32 & 00:36:41.0 & +08:19:10 & 2.224  & 1.303 & 13.8 & 17.65 $\pm$ 0.11  & 1.142 & 1.998 \\ 
        2020-09-06 08:04 & 00:34:28.6 & +07:18:12 & 2.233  & 1.288 & 11.9 & 17.64 $\pm$ 0.06  & 1.179 & 1.839 \\ 
        2020-09-12 09:28 & 00:30:27.4 & +05:37:19 & 2.247  & 1.273 & 8.6  & 17.49 $\pm$ 0.05   & 1.162 & 1.998 \\ 
        2020-09-12 10:21 & 00:30:25.8 & +05:36:41 & 2.247  & 1.273 & 8.6  & 17.53 $\pm$ 0.05   & 1.248 & 1.954 \\
        2020-09-18 08:41 & 00:25:56.9 & +03:51:55 & 2.260  & 1.267 & 5.3  & 17.35 $\pm$ 0.03    & 1.144 & 2.073 \\ 
        2020-09-20 09:39 & 00:24:19.3 & +03:15:05 & 2.264  & 1.268 & 4.1  & 17.30 $\pm$ 0.03    & 1.226 & 1.548 \\ 
        2020-09-21 07:17 & 00:23:36.0 & +02:58:44 & 2.266  & 1.268 & 3.6  & 17.26 $\pm$ 0.03   & 1.164 & 1.785 \\ 
        2020-09-21 07:34 & 00:23:35.4 & +02:58:32 & 2.266  & 1.268 & 3.6  & 17.27 $\pm$ 0.03   & 1.15  & 2.031 \\ 
        2020-09-23 07:54 & 00:21:57.3 & +02:21:58 & 2.271  & 1.270 & 2.4  & 17.17 $\pm$ 0.03    & 1.139 & 1.996 \\ 
        2020-09-26 06:20 & 00:19:33.9 & +01:28:48 & 2.277  & 1.275 & 0.8  & 17.07 $\pm$ 0.03   & 1.22  & 1.520 \\ 
        2020-09-27 09:09 & 00:18:39.1 & +01:08:41 & 2.280  & 1.278 & 0.4  & 16.96 $\pm$ 0.03    & 1.22  & 1.612 \\ 
        2020-10-04 07:07 & 00:13:10.5 & -00:52:54 & 2.295  & 1.302 & 3.8  & 17.39 $\pm$ 0.05   & 1.233 & 1.894 \\ 
        2020-10-06 06:03 & 00:11:42.3 & -01:25:55 & 2.300  & 1.311 & 4.9  & 17.42 $\pm$ 0.05     & 1.265 & 1.621 \\ 
        2020-10-08 06:07 & 00:10:14.9 & -01:58:53 & 2.304  & 1.321 & 6.0  & 17.48 $\pm$ 0.03   & 1.252 & 1.853 \\ 
        2020-10-12 07:07 & 00:07:29.7 & -03:02:29 & 2.313  & 1.346 & 8.0  & 17.62 $\pm$ 0.04   & 1.252 & 2.267 \\ 
        2020-10-14 07:09 & 00:06:14.4 & -03:32:21 & 2.317  & 1.360 & 9.0  & 17.67 $\pm$ 0.03    & 1.262 & 1.848 \\
        \hline
    \caption{Columns: (1) Date of observation; (2) Right Ascension; (3) Declination; (4) Heliocentric distance at time of observation; (5) Geocentric distance at time of observation; (6) Phase Angle at time of observation; (7) Magnitude; (8) Airmass of observation; (9) In-image seeing}
    \label{table4}
\end{longtable}

\begin{figure}[b!]
    \centering
    \includegraphics[width=\linewidth]{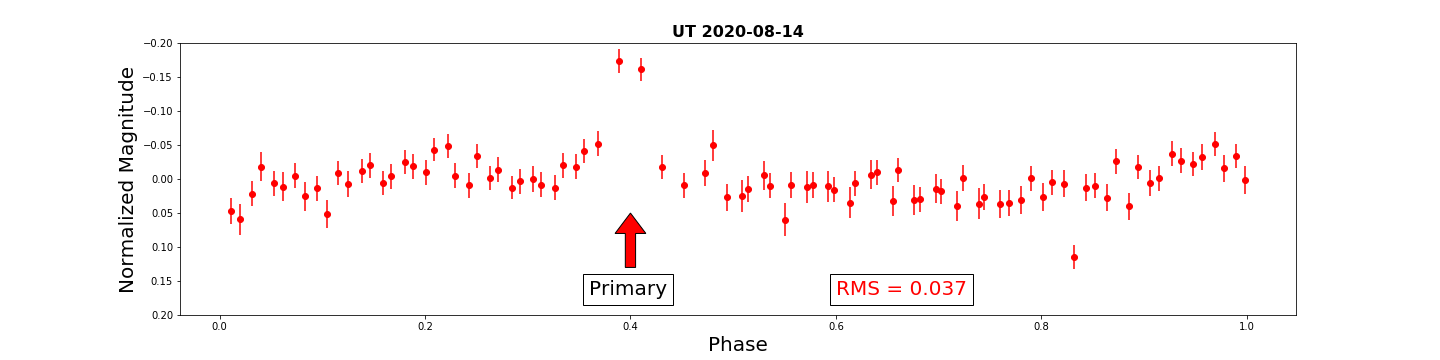}
    \caption{Phase-folded lightcurve of Gault on 2020 August 14 UT from MLO folded by 2.5 hours. Unfortunately, Gault passed over a star in the same observation frame and caused the magnitude to spike more than expected just as it was showing a peak. The secondary peak is somewhat undefined compared to the noise.}
    \label{0814_Phase}
\end{figure}

\begin{figure}
    \centering
    \includegraphics[width=\linewidth]{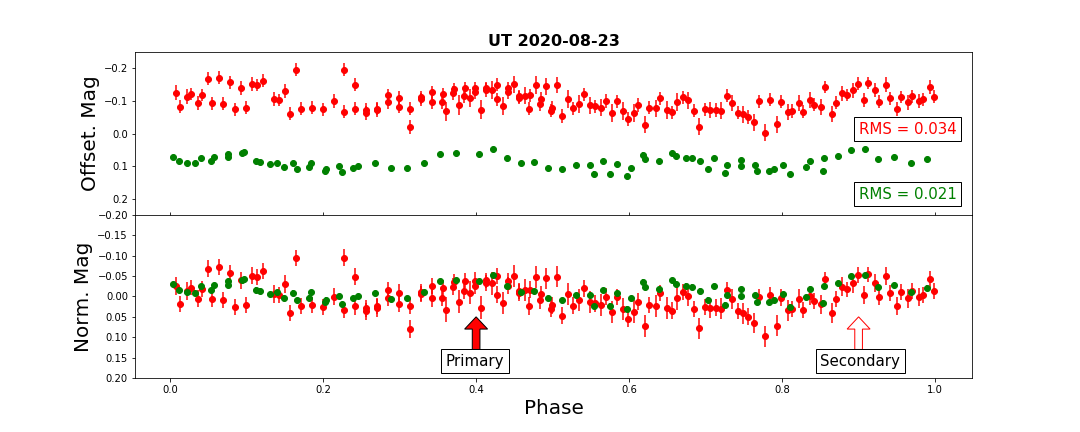}
    \caption{Phase-folded lightcurve of Gault on 2020 August 23 UT from MLO and LOT folded by 2.5 hours. Although the primary peak is not well-defined in the MLO observations, the LOT observations show a promising bump. The secondary peak also is somewhat defined in both lightcurves.}
    \label{0823_Phase}
\end{figure}

\begin{figure}
    \centering
    \includegraphics[width=\linewidth]{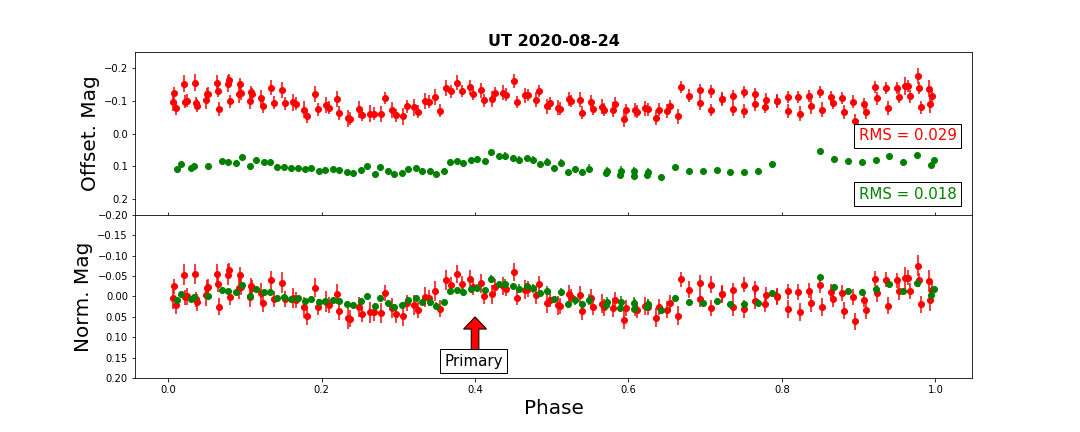}
    \caption{Phase-folded lightcurve of Gault on 2020 August 24 UT from MLO and LOT folded by 2.5 hours. Both lightcurves display a promising primary bump but neither have a convincing secondary bump.}
    \label{0824_Phase}
\end{figure}

\begin{figure}
    \centering
    \includegraphics[width=\linewidth]{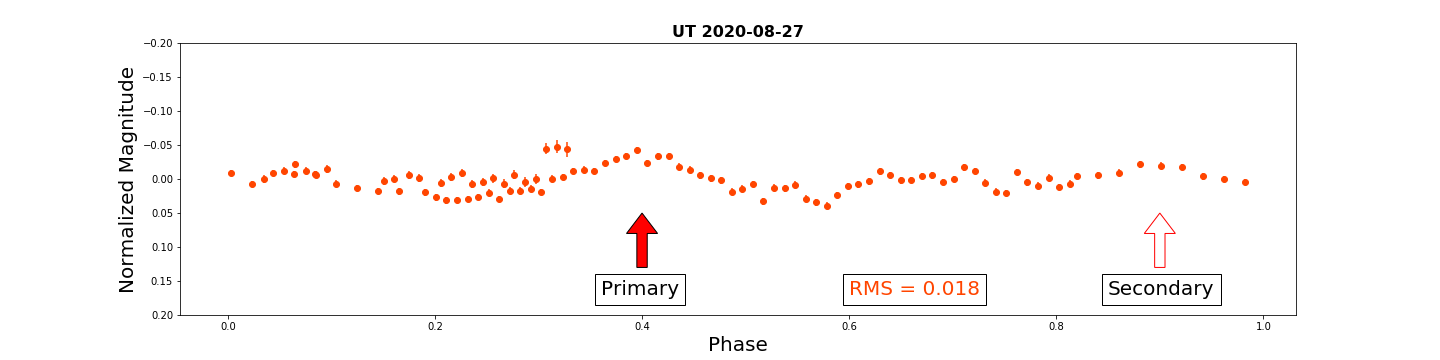}
    \caption{Phase-folded lightcurve of Gault on 2020 August 27 UT from P200 folded by 2.5 hours. Although this lightcurve is short, the high-quality observation displays a strong primary peak and convincing secondary peak.}
    \label{0827_Phase}
\end{figure}

\begin{figure}
    \centering
    \includegraphics[width=\linewidth]{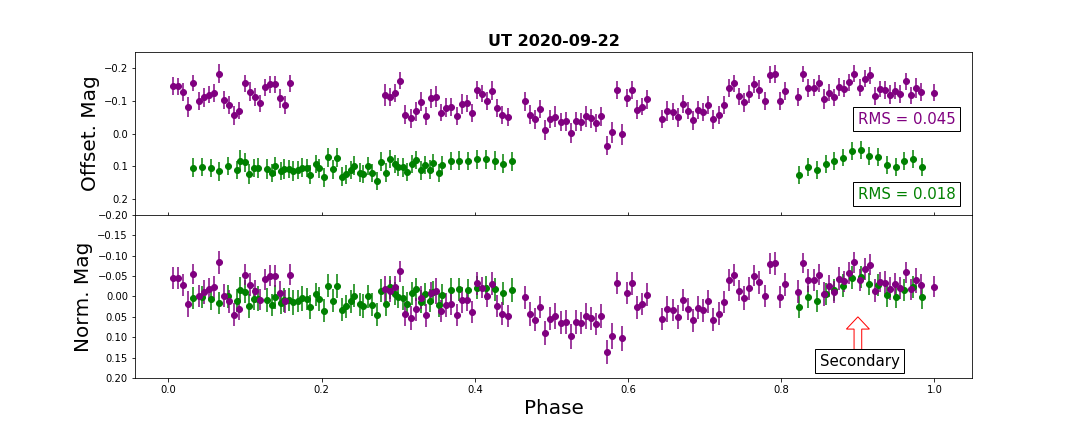}
    \caption{Phase-folded lightcurve of Gault on 2020 September 22 UT from LOT and TMO folded by 2.5 hours. The primary peak does not rise beyond the noise in either observation, but both observations display a promising secondary peak.\footnote{In this case it is important to note that the primary and secondary peaks are both separated in time by 2.5 hours from other primary and secondary peaks in the lightcurve, so therefore one could arbitrarily say that the primary is strong and the secondary is weak for this date.}}
    \label{0922_Phase}
\end{figure}

\begin{figure}
    \centering
    \includegraphics[width=\linewidth]{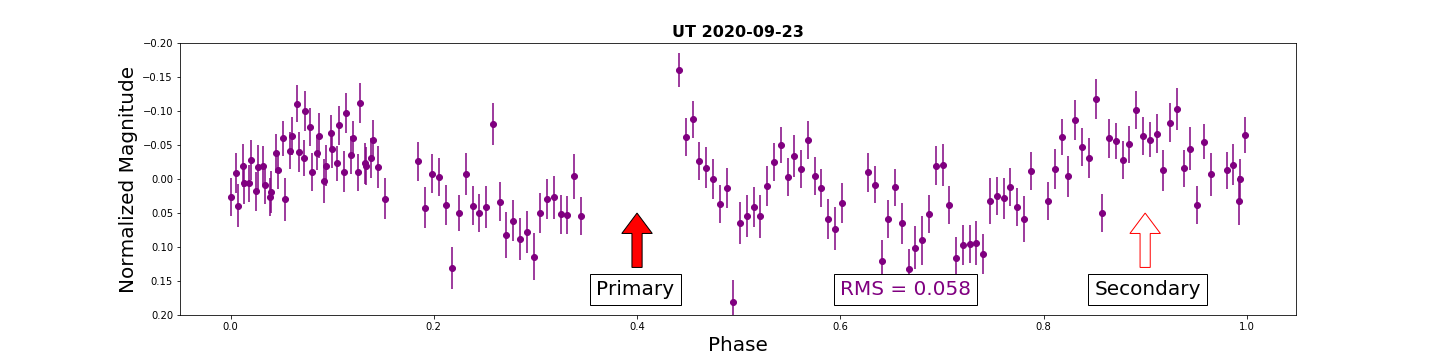}
    \caption{Phase-folded lightcurve of Gault on 2020 September 23 UT from TMO folded by 2.5 hours. Observations from TMO during September of 2020 were affected by a local wildfire and therefore the noise in lightcurves from TMO is abnormally high. The primary looks to have occured just as observations were temporarily halted, but were resumed in time to see a modest secondary peak.}
    \label{0923_Phase}
\end{figure}

\begin{figure}
    \centering
    \includegraphics[width=\linewidth]{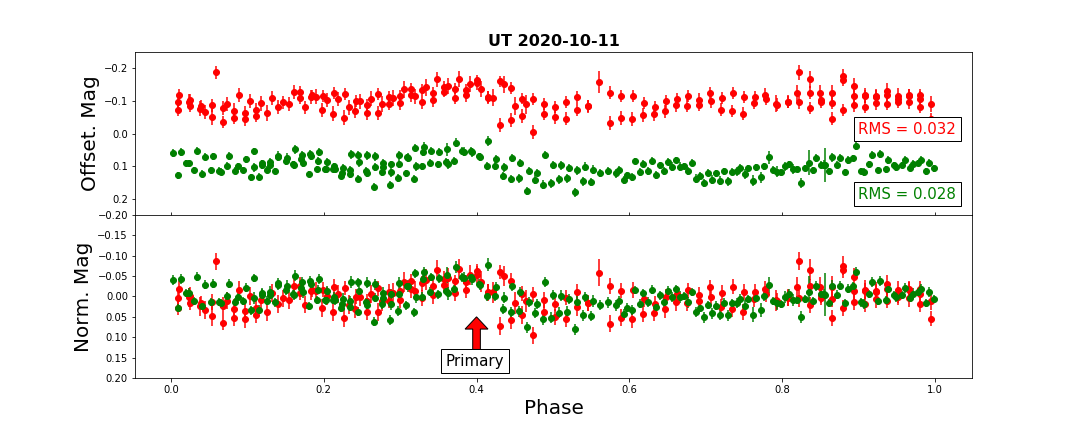}
    \caption{Phase-folded lightcurve of Gault on 2020 October 11 UT from MLO and LOT folded by 2.5 hours. Both lightcurves show a strong primary peak but neither appear to show a strong secondary peak.}
    \label{1011_Phase}
\end{figure}

\begin{figure}
    \centering
    \includegraphics[width=\linewidth]{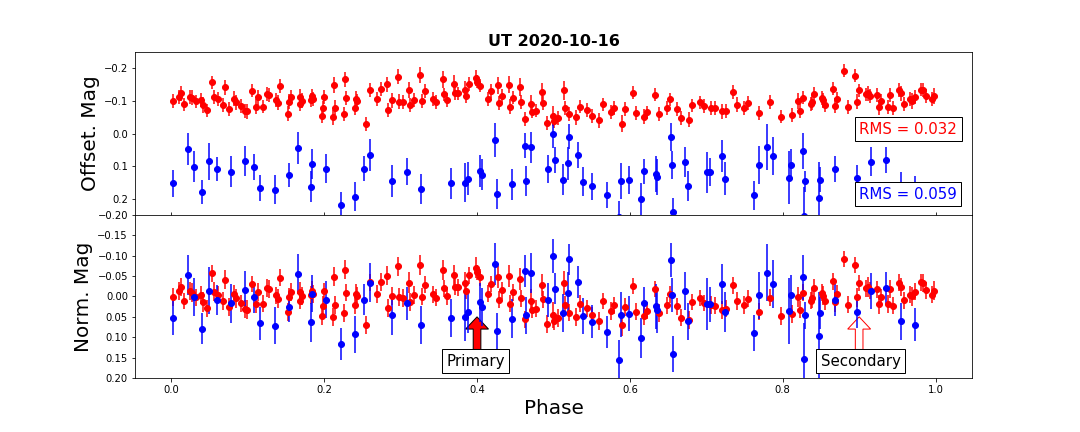}
    \caption{Phase-folded lightcurve of Gault on 2020 October 16 UT from MLO and GIT folded by 2.5 hours. Unfortunately the GIT observations were affected by less-than-ideal seeing on this date (3.1\arcsec; see Table \ref{table2}) and therefore it is difficult to see either peak through the noise. The MLO data show possible primary and secondary peaks but are also affected by noise.}
    \label{1016_Phase}
\end{figure}

\begin{figure}
    \centering
    \includegraphics[width=\linewidth]{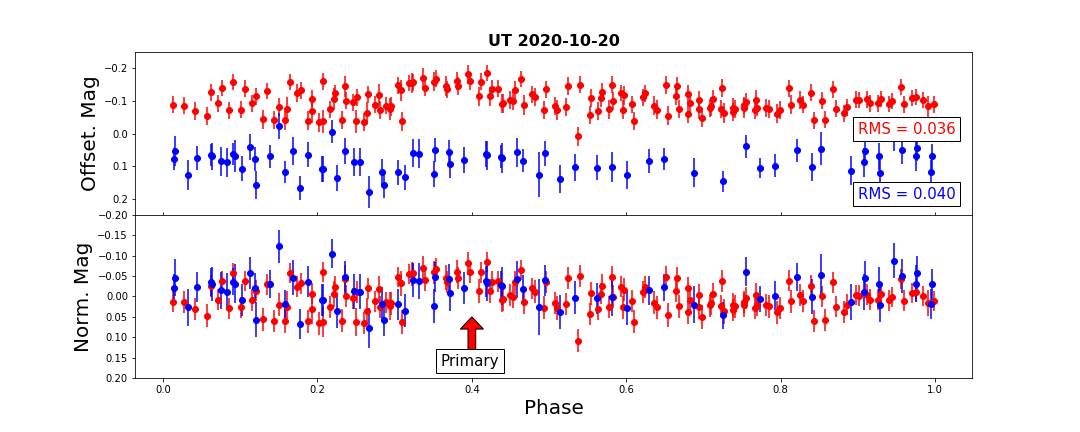}
    \caption{Phase-folded lightcurve of Gault on 2020 October 20 UT from MLO and GIT folded by 2.5 hours. Both lightcurves show promising primary peaks but do not display strong secondary peaks.}
    \label{1020_Phase}
\end{figure}

\end{document}